\documentclass[12pt,a4paper,aps,preprint,superscriptaddress,nofootinbib]{revtex4-1}
\usepackage[utf8]{inputenc}
\usepackage{graphicx}
\usepackage{amssymb}
\usepackage{textcomp}
\usepackage{amsmath}
\usepackage{tabularx}
\usepackage{bm}
\usepackage{times}
\usepackage{color}
\usepackage{slashed}
\usepackage{multirow}
\usepackage{verbatim}
\usepackage{cancel}
\usepackage{subfigure}
\usepackage{ulem}

\usepackage[colorlinks=true, pdfstartview=FitV, linkcolor=blue, citecolor=blue, urlcolor=blue]{hyperref}
\allowdisplaybreaks[4]

\def\lsim{\mathrel{\raise.3ex\hbox{$<$\kern-.75em\lower1ex\hbox{$\sim$}}}}
\def\gsim{\mathrel{\raise.3ex\hbox{$>$\kern-.75em\lower1ex\hbox{$\sim$}}}}

\definecolor{orange}{rgb}{1,0.5,0}

\begin{document}

\title{Axion-like particles from primordial black hole evaporation and their detection in neutrino experiments}

\author{Tong Li}
\email{litong@nankai.edu.cn}
\affiliation{
School of Physics, Nankai University, Tianjin 300071, China
}
\author{Rui-Jia Zhang}
\email{zhangruijia@mail.nankai.edu.cn}
\affiliation{
School of Physics, Nankai University, Tianjin 300071, China
}

\begin{abstract}
The primordial black holes (PBHs) play as a novel source to radiate light elementary particles of energies in the region of a few hundred MeV. We explore the possibility that the light axion-like particles (ALPs) are produced from PBH evaporation. The absorption of light ALPs in the underground detector targets then induces energetic photoelectron signatures in current and future neutrino experiments. Utilizing the PBH ALP event rate, we place general exclusion limits on the axion couplings at Super-K and Hyper-K. We also translate these limits into the upper bound on the fraction of DM composed of PBHs $f_{\rm PBH}$.
\end{abstract}

\maketitle

\section{Introduction}
\label{sec:Intro}

Dark Matter (DM) is one of the most evidential facets of possible new physics beyond the Standard Model (SM). Despite plenty of astrophysical evidence of DM existence in the Universe, however, the constitution and the properties of DM are still mystery. Up to now, there is no conclusive evidence of the weakly interacting
massive particle (WIMP) in DM direct
detection (DD) experiments~\cite{LUX-ZEPLIN:2022qhg,Aprile:2022vux}. Much more effort
has recently been paid to the theoretical hypotheses beyond the WIMP and their detection methods in experiments.
The axion and primordial black hole (PBH) are two of the most well-motivated alternative DM candidates.

Axions as a solution to the strong CP problem~\cite{Baluni:1978rf,Crewther:1979pi,Kim:1979if,Shifman:1979if,Dine:1981rt,Zhitnitsky:1980tq,Baker:2006ts,Pendlebury:2015lrz} receive a wide interest in phenomenology and cosmology. Both the QCD axion~\cite{Peccei:1977hh,Peccei:1977ur,Weinberg:1977ma,Wilczek:1977pj} (for a recent review see Ref.~\cite{DiLuzio:2020wdo}) and axion-like particles (ALPs)~\cite{Kim:1986ax,Kuster:2008zz} can play as cold DM through the misalignment mechanism. The general ALPs span a wide range of the mass and the coupling constants in a variety of theories, from sub-eV wavelike axion to MeV or even heavier point particle.

The PBHs formed from the collapse of local overdensities in the early Universe can serve as  hypothetical macroscopic objects in DM halos~\cite{Zeldovich:1967lct,Carr:1974nx,Carr:1975qj,Khlopov:2008qy} (see Ref.~\cite{Carr:2021bzv} for a recent review).
Hawking's famous discovery tells that PBHs with the mass $\sim 10^{15}$ g would thermally radiate elementary particles~\cite{Hawking:1974rv} and thus they cannot provide all the observed abundance of DM~\cite{Barrau:2003xp}. The emitted particles such as gamma-rays, neutrinos and $e^\pm$ in the evaporation process suffer from a variety of cosmological constraints~\cite{Carr:2020gox,Laha:2019ssq,Laha:2020ivk,Saha:2021pqf,Ray:2021mxu,Dasgupta:2019cae,Wang:2020uvi,Calabrese:2021zfq,DeRomeri:2021xgy,Ghosh:2021vkt,Capanema:2021hnm,Chao:2021orr,Bernal:2022swt}. The elementary DM particles emitted from the present PBHs were also studied in very recent works~\cite{Calabrese:2021src,Li:2022jxo,Calabrese:2022rfa}.
The emitted particles can acquire energies to a few hundred MeV from the evaporation of PBHs with the mass around $10^{14}-10^{15}$ g. The light particles then gain enough kinetic energies to travel through the Earth and reach the terrestrial detectors.

An interesting scenario is that the PBHs play as a novel source to produce light boosted ALPs~\cite{Calza:2021czr,Schiavone:2021imu}. The ALPs reach the Earth and interact with the targets in current and future terrestrial facilities. The absorption of ALPs in the underground detector targets then induces photoelectron signatures in the time-projection
chambers (TPCs).
We refer the light ALPs from the PBH evaporation as ``PBH ALP'' below.
Unlike the solar axion residing mostly in the keV energy region, PBH ALPs possess energies up to $\mathcal{O}(100)$ MeV and are boosted to (semi-)relativistic velocities. As a result, the neutrino scattering experiments are sensitive to the PBH ALP in light of the energetic photoelectron signature. Moreover, the solar axions are produced by the Primakoff process, the ``ABC'' processes and the nuclear transitions governed by the axion-photon coupling, axion-electron coupling and axion-nucleon coupling, respectively. By contrast, the calculation of PBH radiation into elementary particles are based on quantum mechanics in curved spacetime. The conversion between the vacuum quantum states close to the black hole and far away from the black hole leads to a net
thermal flux of particles with an emission rate (see Refs.~\cite{Hawking:1974rv,Carr:1974nx,Carr:1975qj} and recent reviews~\cite{Carr:2020gox,Carr:2021bzv,Auffinger:2022khh}).
Thus, the PBH ALP flux is independent of the axion coupling constants. Only the scattering processes in the terrestrial detectors rely on the couplings and we thus expect the relevant constraints would be stronger in this sense.

In this work we explore the light ALPs from the PBH evaporation in light of the neutrino experiments.
Super-Kamiokande (Super-K) with 161.9 kiloton$\cdot$years exposure~\cite{Super-Kamiokande:2017dch} provides a strong constraint on sub-GeV boosted DM~\cite{Ema:2018bih}. The typical kinetic energy of the recoiled electron is above 100 MeV at Super-K. One of us studied the DM flux from PBH and the consequent DM-electron scattering at Super-K~\cite{Li:2022jxo}. Here we will evaluate the ALP scattering induced by the axio-electric (AE) effect and the inverse Primakoff (IP) process. In addition, a hypothetical axion-dependent nucleon electric dipole moment (nEDM)~\cite{Lucente:2022vuo} is introduced and we also examine the relevant PBH ALP event rate. Utilizing the PBH ALP event rates and the Super-K IV data, we expect to construct a method to place general exclusion limits on the axion couplings.
The next-generation water Cherenkov detector Hyper-Kamiokande (Hyper-K)~\cite{Hyper-Kamiokande:2018ofw}, currently under construction, has larger fiducial volume of the detector and can improve the current sensitivity of T2K. We will also obtain the prospects of axion couplings for PBH ALP in future Hyper-K experiment.
On the other hand, these limits can also be
translated into the upper bound on the fraction of DM composed of PBHs $f_{\rm PBH}$. We expect the bound on $f_{\rm PBH}$ can be improved from neutrino experiments, compared with the current evaporation constraints from extragalactic gamma-rays.

This paper is organized as follows. In Sec.~\ref{sec:ALPflux} we evaluate
the ALP emission rate and flux from PBH evaporation. The detection event rate in the terrestrial facilities
are then calculated in Sec.~\ref{sec:Rate}. We discuss the axio-electric effect, the inverse Primakoff process as well as a hypothetical axion-dependent nucleon dipole portal. The resultant
constraints on axion couplings and $f_{\rm PBH}$ from Super-K and the future prospects in Hyper-K are displayed in Sec.~\ref{sec:Detection}. Our conclusions are drawn in Sec.~\ref{sec:Con}.

\section{ALP flux from PBH evaporation}
\label{sec:ALPflux}

The famous discovery by Hawking shows that the PBHs have quantum properties and thermally radiate with a temperature $T_{\rm PBH}$ given by~\cite{Hawking:1975vcx,Page:1976df,Page:1977um,MacGibbon:1990zk,MacGibbon:1991tj}
\begin{eqnarray}
k_{\rm B}T_{\rm PBH}={\hbar c^3\over 8\pi G M_{\rm PBH}}\approx 1.06\Big({10^{16}~{\rm g}\over M_{\rm PBH}}\Big)~{\rm MeV}\;,
\end{eqnarray}
where $k_{\rm B}$ is the Boltzmann constant, $G$ is the Newtonian constant of gravitation, and $M_{\rm PBH}$ denotes the PBH mass.
New particles of a mass less than $T_{\rm PBH}$ can be emitted during the PBH evaporation. The PBHs with $M_{\rm PBH}\lesssim 10^{16}$ g would produce elementary particles lighter than 1 MeV which can have ultra-relativistic velocities.
The general emission rate can be written as~\cite{Hawking:1971ei,Page:1976df,Page:1976ki}
\begin{equation}
\frac{d^2N}{dEdt}=\frac{g}{2\pi}\frac{\Gamma(E,M_{\rm PBH},a^\ast)}{e^{E/T_{\rm PBH}}-(-1)^{2s}}\;,
\end{equation}
where $E$ is the energy of emitted particle, $a^\ast$ is the spin of PBH, $\Gamma$ is the so-called ``graybody'' factor describing the probability of elementary particles escaping the PBH gravitational well, and $g$ and $s$ are the degrees of freedom and spin of the particle, respectively. Here we consider the ALP emitted from PBH through the Hawking radiation and thus set $g=1$, $s = 0$. In principle, spinning PBHs evaporate faster than the non-spinning PBHs and thus contribute to the content of elementary particles nowadays for $M_{\rm PBH}\gtrsim 7\times 10^{15}~{\rm g}$~\cite{Dasgupta:2019cae}.
We use the public code BlackHawk v2.0~\cite{Arbey:2019mbc,Arbey:2021mbl} to calculate
the above number of ALP per units time and energy emitted by PBHs~\cite{Schiavone:2021imu}.~\footnote{BlackHawk v2.0 provides the additional emission of massive particle and we
directly apply it to our massive ALP case.}

The ALP flux emitted from PBH can then be divided into the
contributions of both the PBHs in galactic halo and the extragalactic PBHs
\begin{equation}
\frac{d^2\varphi_a}{dEd\Omega}=\frac{d^2\varphi_a^{\rm{gal}}}{dEd\Omega}+\frac{d^2\varphi_a^{\rm{egal}}}{dEd\Omega}\;,
\end{equation}
where $\varphi_a^{\rm gal}$ and $\varphi_a^{\rm egal}$ correspond to the ALP flux from galactic (gal) PBHs and extragalactic (egal) PBHs, respectively, and $\Omega$ is the solid angle.
The differential galactic ALP flux is given by
\begin{eqnarray}
{d^2\varphi_a^{\rm gal}\over dE d\Omega}={f_{\rm PBH}\over 4\pi M_{\rm PBH}} {d^2N\over dE dt} \int {d\Omega_s\over 4\pi}\int dl \rho_{\rm MW}[r(l,\psi)]\;,
\label{eq:MWnuflux}
\end{eqnarray}
where $f_{\rm PBH}$ is the fraction of DM composed of PBHs, $\rho_{\rm MW}[r(l,\psi)]$ denotes the DM density in the Milky Way (MW) halo, $r(l,\psi)=\sqrt{r_\odot^2+l^2-2lr_\odot \cos\psi}$ is the galactocentric distance with $r_\odot$ the solar distance from the galactic center, $l$ the line-of-sight distance to the PBH and $\psi$ the angle between these two directions. The maximum limit of distance $l$ is taken to be $(r_h^2-r_\odot^2 \sin^2\psi)^{1/2}+r_\odot \cos\psi$ with the halo radius $r_h=200$ kpc~\cite{Wang:2020uvi}. The angular integration is defined as $\int d\Omega_s=\int_0^{2\pi}d\phi \int_0^\pi d\psi \sin\psi$ with the azimuthal angle $\phi$. We adopt the generalized Navarro-Frenk-White (NFW) DM profile~\cite{Navarro:1996gj}
\begin{eqnarray}
\rho_{\rm MW}(r)=\rho_\odot \Big({r\over r_\odot}\Big)^{-\gamma} \Big({1+r_\odot/r_s\over 1+r/r_s}\Big)^{3-\gamma}\;,
\end{eqnarray}
where $\rho_\odot=0.4~{\rm GeV}/{\rm cm}^3$ is the local DM density~\cite{Salucci:2010qr}, $r_\odot=8.5$ kpc is the distance between the Sun and the Galactic center~\cite{Yuan:2017ozr}, $r_s=20$ kpc is the radius of the galactic diffusion disk, and the inner slope of the NFW halo profile is fixed as $\gamma=1$. For the extragalactic contribution, the differential ALP flux over the full sky is
\begin{eqnarray}
{d^2\varphi_a^{\rm egal}\over dE d\Omega}={f_{\rm PBH}\rho_{\rm DM}\over 4\pi M_{\rm PBH}}\int^{t_{\rm max}}_{t_{\rm min}} dt [1+z(t)]{d^2N\over dE dt}\Big|_{E_s}\;,
\label{eq:EGnuflux}
\end{eqnarray}
where the average DM density of the Universe at the
present epoch is $\rho_{\rm DM}=2.35\times 10^{-30}~{\rm g}/{\rm cm}^3$ determined by Planck~\cite{Planck:2018vyg}, and $E_s=\sqrt{(E^2-m_a^2)(1+z(t))^2+m_a^2}$ denotes the energy at the source which is related to the energy $E$ in the observer's frame by the redshift $z(t)$ for massive ALP. In the massless limit, it simplifies to $E_s=E(1+z(t))$. The ALP emitted from the PBHs in the very early Universe is sufficiently redshifted.
As a result, their energies are very low today and cannot be detected. Thus, for the integral limits, we fix $t_{\rm min}=10^{11}$ s being close to the era of matter-radiation equality~\cite{Wang:2020uvi} and $t_{\rm max}$ as the age of Universe. In fact, it turns out that changing the lower limit has no impact on the results~\cite{DeRomeri:2021xgy}.

\begin{figure}[htb!]
\begin{center}
\includegraphics[scale=1,width=0.47\linewidth]{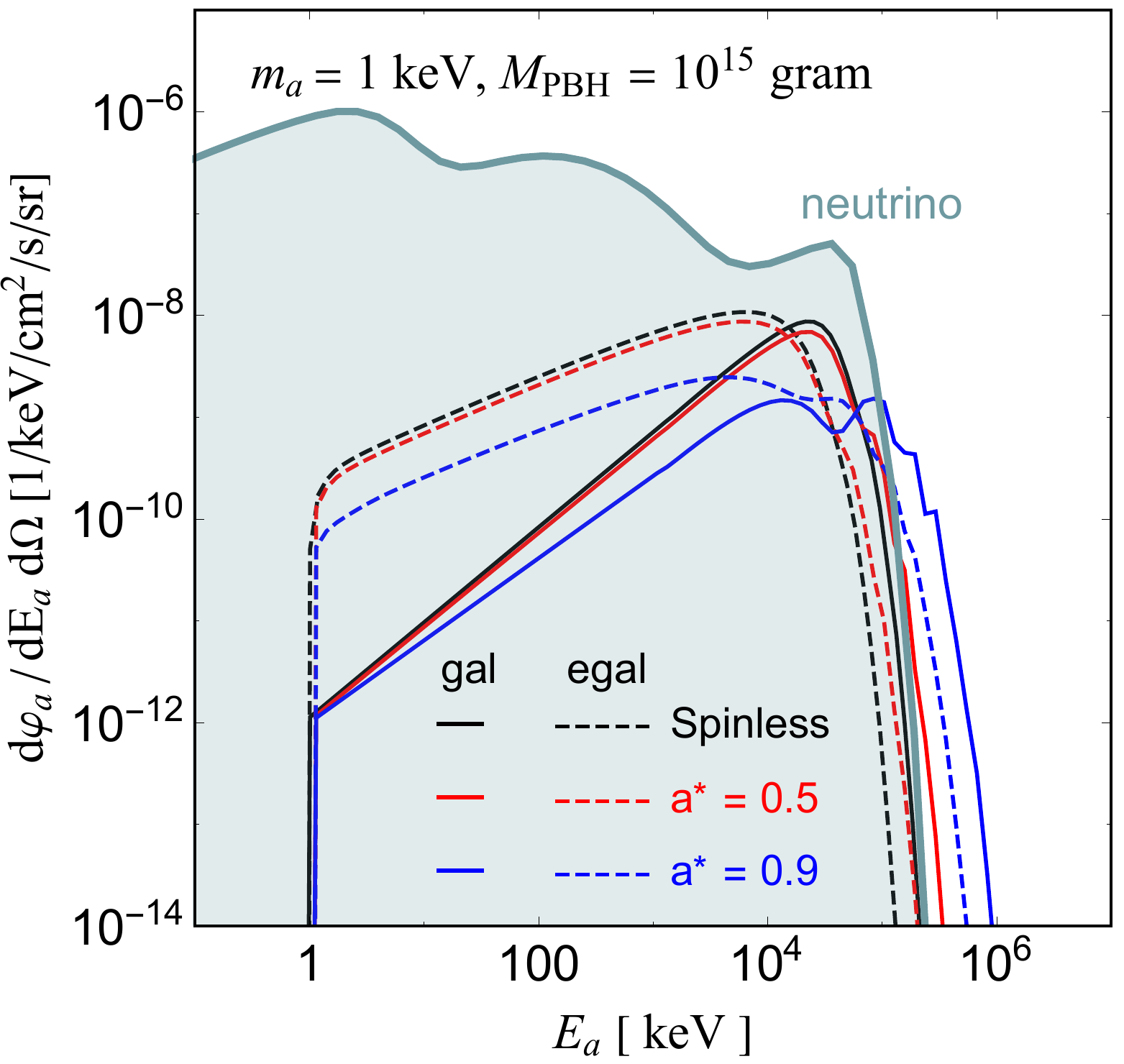}
\includegraphics[scale=1,width=0.47\linewidth]{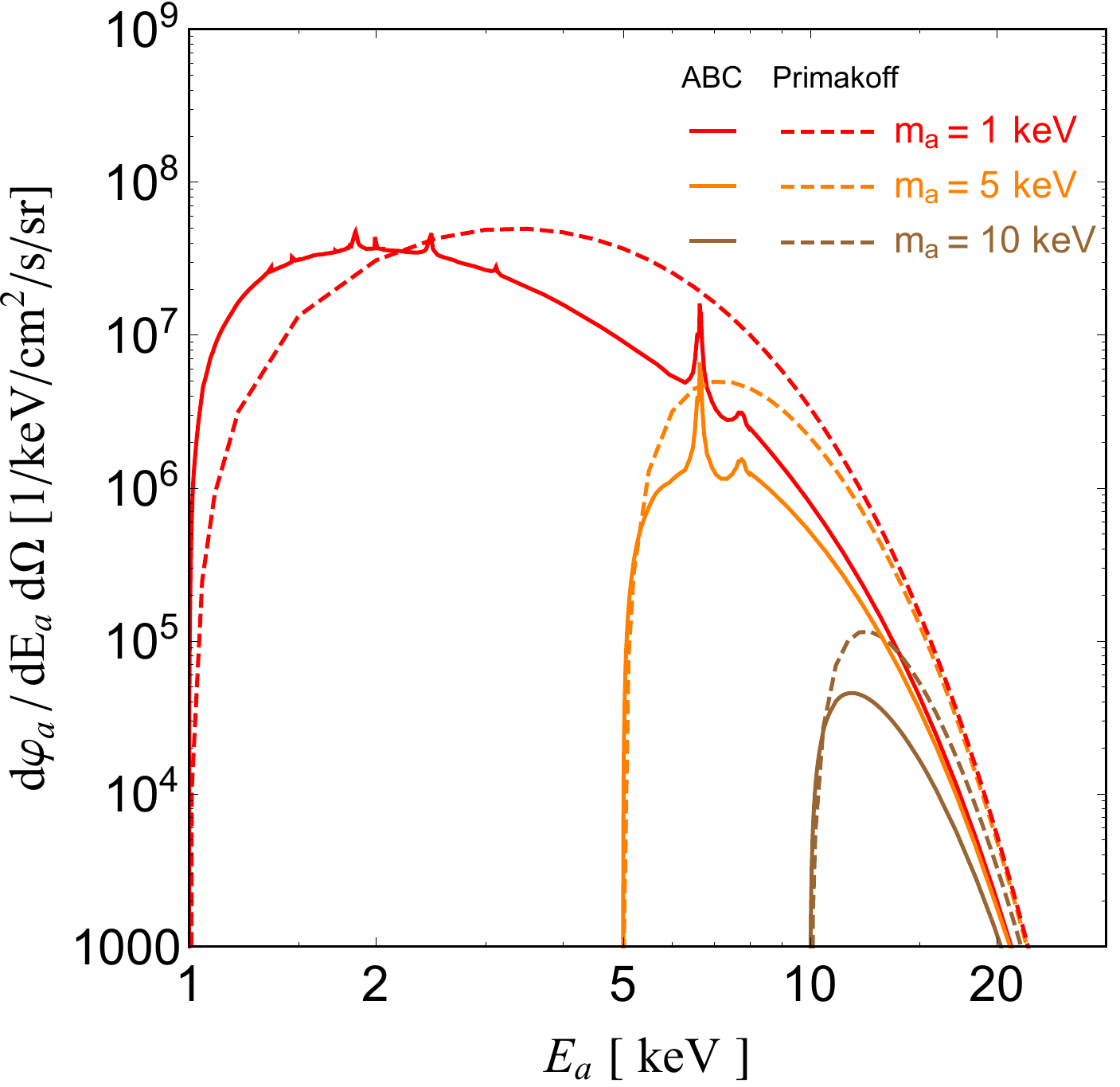}
\end{center}
\caption{Left: The ALP flux from PBH evaporation, for different PBH spins (spinless: black line, $a^\ast=0.5$: red line and $a^\ast=0.9$: blue line) with fixed ALP mass $m_a=1$~keV, PBH mass $M_{\rm{PBH}}=10^{15}$~g and $f_{\rm PBH}=3.9\times 10^{-7}$.
The galactic and extragalactic contributions are denoted by solid and dashed lines, respectively. The neutrino flux is also shown for comparison (suppose spinless PBH with the same mass and $f_{\rm PBH}$). Right: The flux of solar axion, which consists of Primakoff process (dashed) and ABC production (solid) with the coupling constants $g_{a\gamma}=10^{-11}~ \rm{GeV}^{-1}$ and $g_{ae}=10^{-13}$. We set three finite values of axion mass: $m_a=$~1 keV (red), 5 keV (orange) and 10 keV (brown).
}
\label{fig:ALP_flux}
\end{figure}

In the left panel of Fig.~\ref{fig:ALP_flux}, we show the differential ALP flux from PBHs $d^2\varphi_a/dE d\Omega$ for $m_a=1$ keV, $M_{\rm PBH}=10^{15}$ g and $f_{\rm PBH}=3.9\times 10^{-7}$ allowed by extragalactic gammay-ray constraints~\cite{Carr:2020gox}. The galactic and extragalactic contributions are denoted by solid and dashed lines, respectively. One can see that the ALP produced from PBH gains energies peaked at order of $\mathcal{O}(10)-\mathcal{O}(100)$ MeV. The light ALPs would thus receive ultra-relativistic velocities and are highly boosted. As seen in Eqs.~(\ref{eq:MWnuflux}) and (\ref{eq:EGnuflux}), smaller PBHs would exhibit harder spectra of the emitted ALP. We also demonstrate that spinning PBHs make the ALPs gain higher energies along with increasing PBH spin, because the black hole with non-zero angular moment can accelerate the Hawking evaporation. In the following study, as being customary in the literature~\cite{Wang:2020uvi,Calabrese:2021zfq,Calabrese:2021src}, we focus on $M_{\rm PBH}<1\times 10^{16}$ g and only consider non-spinning PBHs.
To compare with the ALP, we also display the neutrino flux from PBH evaporation. There are two contributions to the neutrino emission rate generated by the Hawking radiation~\cite{Dasgupta:2019cae,Wang:2020uvi,Calabrese:2021zfq,DeRomeri:2021xgy,Ghosh:2021vkt,Capanema:2021hnm,Chao:2021orr,Bernal:2022swt}. One of them is the primary contribution consisting of neutrinos directly emitted in the evaporation. The other one is the secondary contribution which origins from the hadronization and the subsequent decay of the primary particles. It turns out that the inclusion of secondary production injects more neutrinos at low energies.

To further demonstrate the properties of ALPs from PBH evaporation, we need to compare with the conventional source of axions produced in the Sun.
The components of solar axion mainly originate from the following two mechanisms: (i) the Primakoff production process (i.e. $\gamma \mathbb{Z}\to a \mathbb{Z}$ with $\mathbb{Z}$ denoting the atomic system $\mathbb{Z}=(e^-,N)$) which depends on the axion-photon coupling $g_{a\gamma}$, and (ii) the so-called ``ABC'' process~\cite{Redondo:2013wwa}. The ``ABC'' production consists of three processes: the Atomic de-excitation and recombination, Bremsstrahlung, and Compton scattering process. These production channels are all relevant with axion-electron coupling $g_{ae}$ via the axion-lepton interaction term.
Thus, the relevant interaction terms of solar axion are
\begin{equation}
\mathcal{L}\supset-\frac{1}{4}g_{a\gamma}aF_{\mu\nu}\tilde{F}^{\mu\nu}+ig_{ae}a\bar{\psi}_e\gamma_5\psi_e\;.
\end{equation}
In previous studies of solar axion flux, people usually utilized the property of axion being a nearly massless pseudo-scalar boson whose mass is far less than $\mathcal{O}(1)$ eV. Thus, the calculations of solar axion flux can be simplified under the ultra-relativistic approximation. Nevertheless, computing the flux of ALP with finite mass requires to generally extend the method. In more detail, we should recompute the matrix element and phase space integral without the approximation of massless axion. In fact, some computations of solar axion flux depending on arbitrary $m_a$ have been done in previous works~\cite{Raffelt:1985nk,Gondolo:2008dd,Caputo:2020quz,Carenza:2021osu}.
We follow them to explicitly calculate the primakoff, bremsstrahlung and compton scattering process.
For the flux produced in axio-deexcitation and axio-recombination (i.e. free-bound and bound-bound electron transitions), we simply use the data in Ref.~\cite{Redondo:2013wwa} and interpolate them to energies above 10 keV.
We also multiply every flux component (even though it is not rigorous for ``A'' flux) by a rescaling factor $(1-m_a^2/E_a^2)$ to compensate the variation from phase space integrals of energy-loss rate for non-zero $m_a$ case.

The solar ALP flux is shown in the right panel of Fig.~\ref{fig:ALP_flux} for different choices of $m_a=1, 5, 10$ keV. It turns out that the PBH ALP flux is nearly 16 orders of magnitude smaller than the solar axion with the coupling constants $g_{a\gamma}=10^{-11}~{\rm GeV}^{-1}$ and $g_{ae}=10^{-13}$ favored by the stellar cooling~\cite{Giannotti:2017hny}.
However, the spectrum of solar axion drops rapidly after the peak right above the axion mass threshold for $m_a\sim \mathcal{O}(1)$ keV. Thus, the solar axion is usually used to account for the measurements from DM direct detection experiments with low recoil energy, such as the former XENON1T excess in the range of 1-5 keV~\cite{XENON:2020rca}. By contrast, the ALP flux from PBH gets enhanced along with increasing energies and peaks well above the threshold around $\mathcal{O}(10)-\mathcal{O}(100)$ MeV.
In spite of much low flux, the PBHs emit more energetic ALPs. The neutrino experiments such as Super-K and Hyper-K are sensitive to MeV-scale energy regime and can provide compelling bounds for ALP from PBH evaporation. We will discuss the event rate in these neutrino experiments and the bounds on axion couplings and PBHs in the following sections.

\section{Event rates at terrestrial detectors}
\label{sec:Rate}

After being captured by detector, there are two main types of scattering
between the target atom and the ALP:
(i) axio-electric effect, referring to the process that ALP is absorbed by bound-state electron of target atom and leads to an ionization signal, which depends on the axion-electron coupling $g_{ae}$, and (ii) inverse Primakoff process $a+\mathbb{Z}\to \gamma+\mathbb{Z}$. In the latter case (ii), one ALP can be converted into two photons through the axion-photon coupling $g_{a\gamma}$. Then, one virtual photon makes atom gain energy (or be excited) while the other is emitted in the form of real photon. In addition, we will also consider the detection channel $a + p \to p +\gamma$ on water Cherenkov detector through the nucleon dipole portal~\cite{Lucente:2022vuo}, which also produces a visible photon signal.
Next, we individually discuss the scattering cross-sections and event rates corresponding to these types of process.

\subsection{The axio-electric effect}

\begin{figure}
\centering
\includegraphics[scale=0.8]{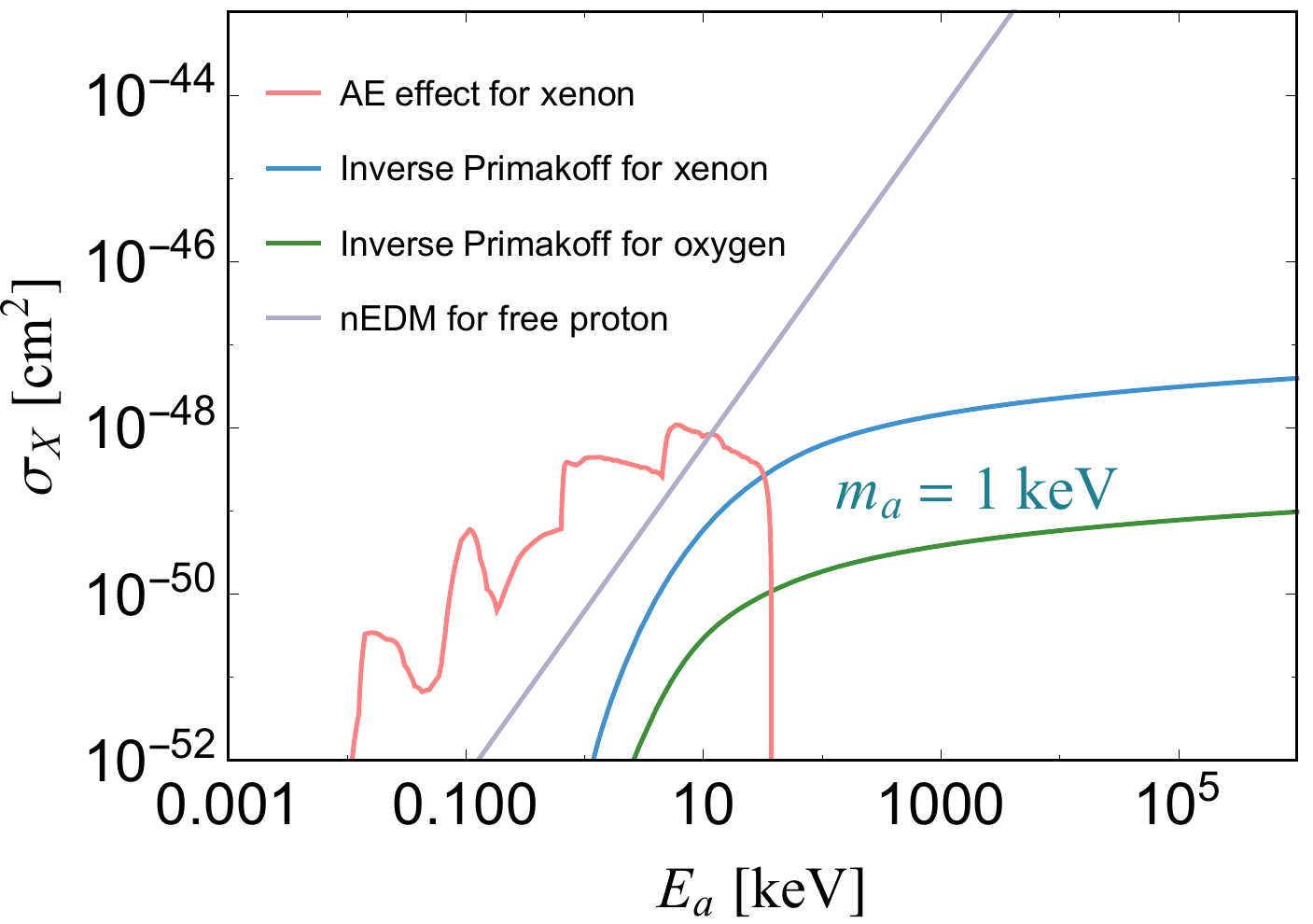}
\caption{Cross-sections for different detection processes including AE effect for $^{131}{\rm Xe}$ (red), inverse Primakoff for xenon (blue), inverse Primakoff for oxygen (green), and nucleon EDM for free proton (purple).
The axion mass is assumed to 1 keV for inverse Primakoff process (only inverse Primakoff process dependent on axion mass). The axion coupling constants are set as $g_{a\gamma}=10^{-11}~\rm{GeV}^{-1}$, $g_{ae}= 10^{-13}$ favored by the stellar cooling~\cite{Giannotti:2017hny,DiLuzio:2020jjp} and $g_d=10^{-11}~\rm{GeV}^{-2}$, respectively.}
\label{fig:cro-section}
\end{figure}

For the AE effect, the total cross-section is given by~\cite{Pospelov:2008jk,Bloch:2016sjj}
\begin{equation}
\sigma_{AE}(E_a)=\sigma_{PE}(E_a)\frac{3g_{ae}^2}{16\pi\alpha_{EM}v_a}\frac{E_a^2}{m_e^2}(1-\frac{1}{3}v_a^{2/3})\;,
\label{equ:sigma_AE}
\end{equation}
where $\alpha_{EM}$ is the fine-structure constant, $v_a$ is the axion velocity and $\sigma_{PE}$ is the photoabsorption cross-section of atoms measured by low-energy x-ray experiment~\cite{Henke:1993eda}. The cross-section gives a narrow jagged peak which helps to separate signal from background as mentioned in Ref.~\cite{Bloch:2016sjj}. In Fig.~\ref{fig:cro-section}, we show $\sigma_{AE}$ for $^{131}{\rm Xe}$ with the axion-electric coupling $g_{ae}=10^{-13}$. The ALP absorption cross-section $\sigma_{AE}$ is limited within the energy range of $0.01\sim 30$ keV because $\sigma_{PE}$ is suppressed by high photon energies. Here we assume that the incoming velocity of axion should be relativistic, as the axion radiated from PBH has a velocity near the speed of light, i.e., $v_a\approx c$. Taking into account the ALP flux $d\varphi_a/dE_a$ as well as the cross-section $\sigma_{AE}$, we can further obtain the differential event rate of the scattering process
\begin{equation}
\frac{dR}{dE_R}={N_A\over m_A}\frac{d\varphi_a}{dE_a}\sigma_{AE}(E_a)\;,
\label{equ:eventrate}
\end{equation}
where $N_A$ is the Avogadro constant and $m_A$ is the molar mass of atom or molecule.
Since the axion mass is much smaller than that of electron or nucleus, the recoil energy delivered by axion is approximately equal to the incoming axion energy $E_a$. Thus, we can directly rewrite the right-handed side of above formula which depends on $E_a$ into the form relying on the recoil energy $E_R$. Furthermore, to compare with the results in Refs.~\cite{XENON:2020rca,Gao:2020wer,Dent:2020jhf,Bloch:2020uzh}, the above event rate $dR/dE_R$ is smeared by a Gaussian distribution with the variance of energy as $\sigma_E~[{\rm keV}]=0.3171~[{\rm keV}^{1/2}]\sqrt{E_R}+0.0015E_R~[{\rm keV}]$~\cite{Bloch:2020uzh} which depends on the detector energy and energy resolution.

\begin{figure}[htb!]
\begin{center}
\includegraphics[scale=1,width=0.47\linewidth]{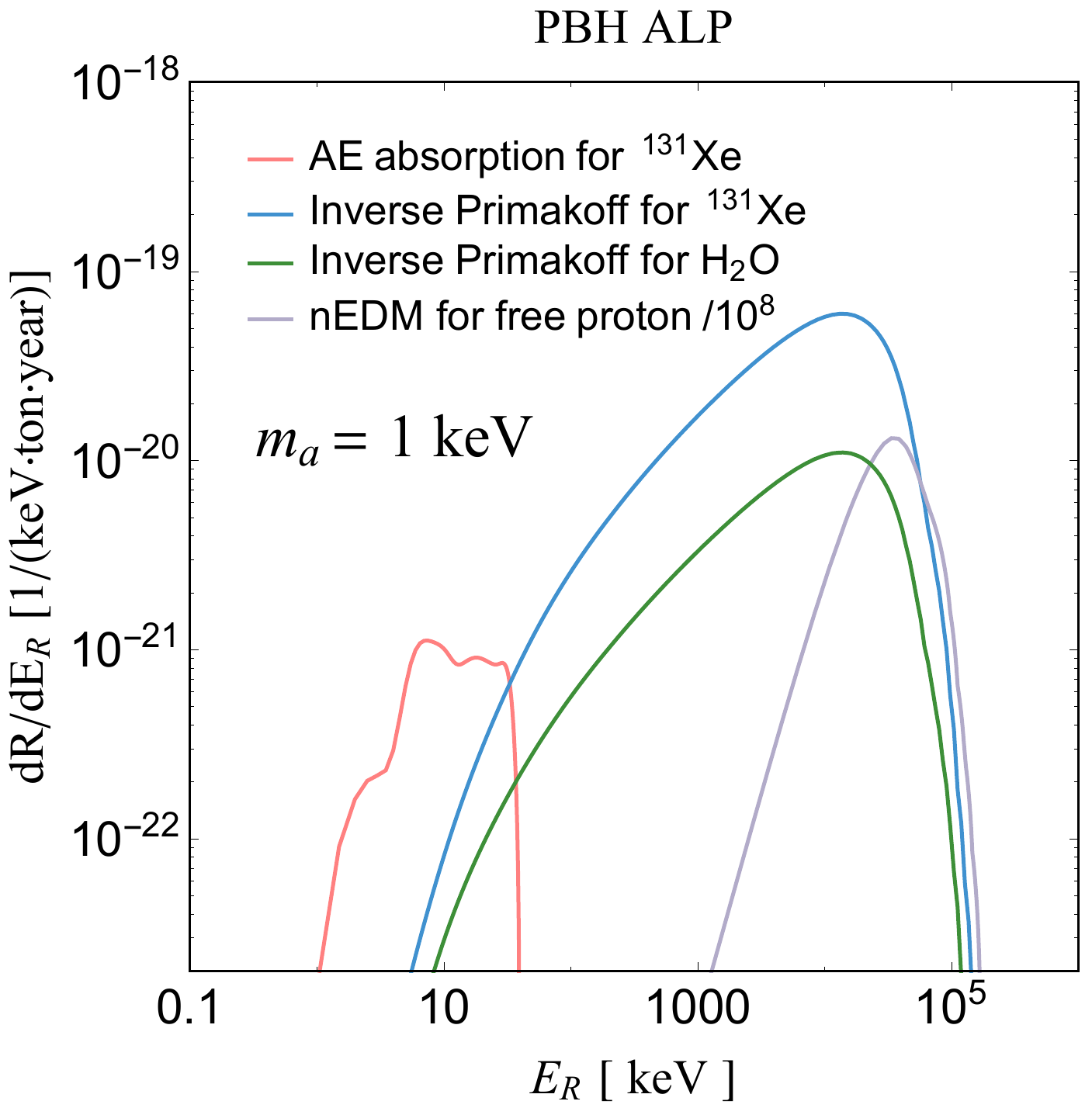}
\includegraphics[scale=1,width=0.47\linewidth]{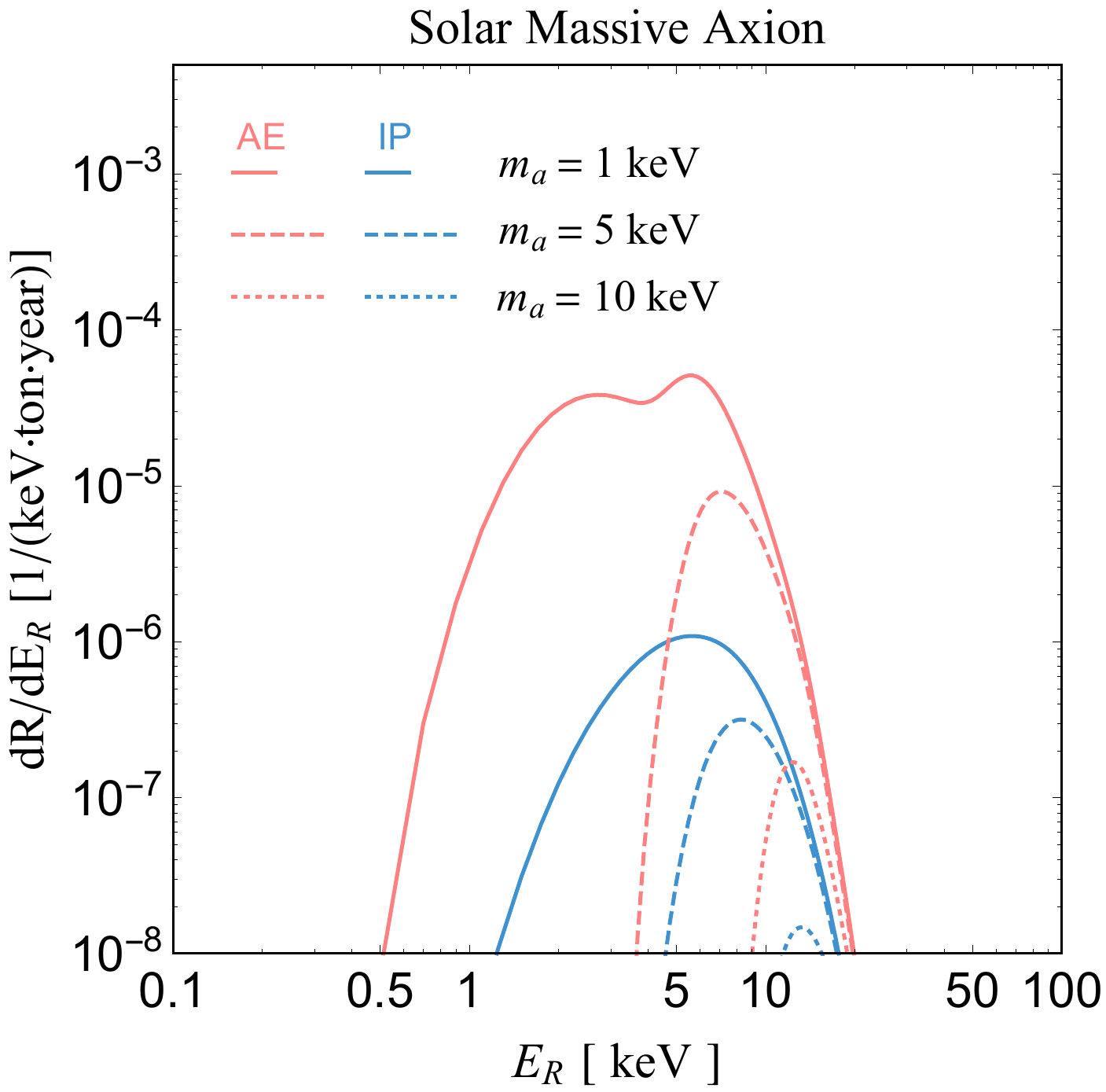}
\end{center}
\caption{Left: Event rates of PBH ALP from AE absorption for $^{131}{\rm Xe}$ (red), inverse Primakoff for $^{131}{\rm Xe}$ (blue), inverse Primakoff for ${\rm H_2O}$ (green) and the nucleon EDM for free proton (purple).
Note that the event rate from nEDM have been scaled by a factor $10^{-8}$ to make them visible. The coupling constants are set as $g_{a\gamma}=10^{-11}~\rm{GeV}^{-1}$, $g_{ae}= 10^{-13}$ and $g_d=10^{-11}~\rm{GeV}^{-2}$, respectively.
We take $M_{\rm PBH}=10^{15}$ g, $f_{\rm PBH}=3.9\times 10^{-7}$ and $m_a=1$ keV. Right: Event rates of solar axion from AE absorption (red) and inverse Primakoff (blue) for $^{131}{\rm Xe}$, in which the solar axion flux is obtained by summing over ABC flux and Primakoff flux. Three different axion masses are assumed as $m_a=1$ keV (solid), 5 keV (dashed) and 10 keV (dotted).
}
\label{fig:EventRate}
\end{figure}

Combining Eq.~(\ref{equ:sigma_AE}) and Eq.~(\ref{equ:eventrate}), we can get the event rate of AE effect as shown in Fig.~\ref{fig:EventRate}. The left (right) panel is for PBH ALP (solar axion). For PBH we take $M_{\rm PBH}=10^{15}$ g and $f_{\rm PBH}=3.9\times 10^{-7}$ for illustration. As a common detection channel for solar axion, due to the sensitive recoil energy below $\mathcal{O}(10)$ keV, AE effect is usually considered to account for the results from DM direct detection. But for the PBH ALP that we focus in this paper, the ALP flux is much small in the working range of AE effect and its contribution is negligible when fitting to the results from XENON1T/XENONnT~\cite{XENON:2020rca,Aprile:2022vux}. In addition, the AE effect loses viability at high energies above several tens of keV. To explore the PBH ALP in the energy range above 100 MeV, we quest for some other detection channels viable at higher energy below.

\subsection{The inverse Primakoff process}

In this subsection, we consider the inverse Primakoff process $a+\mathbb{Z}\to \gamma+\mathbb{Z}$. The outgoing photon would mimic electronic signal as the current detectors are not able to distinguish the signal caused by scattered photons from the electron recoil. Its contribution was taken into account to accommodate the electron recoil events in XENON1T/nT~\cite{Gao:2020wer,Dent:2020jhf,Aprile:2022vux}.

Since the typical mass of the incoming ALP we consider is less than $\sim\rm{MeV}$, the energy transfer is not enough to excite an atom. Thus, we assume that the atom target does not change its state after being scattered, i.e., only the elastic inverse Primakoff process is considered. The differential elastic cross-section per solid angle for $m_a\ll E_a\approx E_\gamma$ is given by~\cite{Abe:2020mcs}
\begin{equation}
\frac{d\sigma_{\rm{el}}}{d\Omega}\simeq \frac{\alpha_{EM} g_{a\gamma}^2\sin\theta}{16\pi(1-\cos\theta)^2}\vert Z-F(\boldsymbol q)\vert^2\;,
\end{equation}
where $Z$ is the atomic number, and $\theta$ is the angle between the momentum of axion and outgoing photon. The atomic form factor $F(\boldsymbol q)$ contains the information about screened effect from electric charge of nucleus. In the previous studies, one usually used the cross-section induced by the screened Coulomb potential: $F(q,r_0)=Z/(1+q^2r_0^2)$~\cite{Abe:2020mcs}, where $r_0$ is the screening length. It gives the cross-section of inverse Primakoff as follows~\cite{Creswick:1997pg,Dent:2020jhf}
\begin{equation}
\sigma_{IP}=\frac{\alpha_{EM} Z^2 g_{a\gamma}^2}{2}\left[\frac{2r_0^2 E_a^2+1}{4r_0^2 E_a^2}\ln{(1+4r_0^2E_a^2)}-1\right]\;,
\end{equation}
where $E_a \approx p_a$ in the massless limit of axion. However, the above formula is only valid for ideal atom. In fact, the most commonly used method in numerical calculation is relativistic-Hartree-Fock (RHF) approximation. Here we take the total cross-section from Ref.~\cite{Abe:2020mcs}
\begin{equation}
\sigma_{IP}^{\rm RHF}=\frac{\alpha_{EM} g_{a\gamma}^2}{8}\int_{E_a-p_a}^{E_a+p_a}{dq\frac{[q^2-(E_a-p_a)^2][(E_a+p_a)^2-q^2]}{p_a^2q^3}\vert Z-F(q)\vert^2}\;,
\end{equation}
where $F(q)$ is the atomic form factor approximated in a simple analytic form, which agrees well with the RHF approximation and improves the efficiency of calculation. Apparently, this cross-section is quadratically dependent on $g_{a\gamma}$. We display the cross-sections for xenon and oxygen atoms in Fig.~\ref{fig:cro-section} with the coupling $g_{a\gamma}=10^{-11}~\rm{GeV}^{-1}$. It turns out that the inverse Primakoff cross-sections get enhanced for large energies of axion.
Replacing $\sigma_{AE}$ in Eq.~(\ref{equ:eventrate}) by $\sigma_{IP}^{\rm RHF}$, one can obtain the event rate of inverse Primakoff process as shown in the left panel of Fig.~\ref{fig:EventRate}.
We can see that the event rate of inverse Primakoff from PBH ALP (blue and green lines) extends to energies as large as 100 MeV and is 1-2 orders of magnitude greater than that of AE absorption due to the energy distribution of ALP flux from PBH evaporation.

For solar axion, we show the event rate of both AE effect and inverse Primakoff process for xenon in the right panel of Fig.~\ref{fig:EventRate}. The energy range of event rate is limited within $1\sim 20$ keV due to the profile of solar axion flux. The event rate for these two detection channels decreases as the axion mass increases.
However, for the PBH ALP in the left panel of Fig.~\ref{fig:EventRate}, we find that there is obvious separation between these two kinds of processes. As we mentioned before, AE effect mainly occurs in low-energy region while at high energies the inverse Primakoff process dominates and starts to drop around $\mathcal{O}(10^4)$ keV. There is a small overlap for the two processes only at the energy of $\mathcal{O}(10)$ keV. It is thus not difficult to distinguish the signals caused by these two types of scatterings. Moreover, it should be emphasized that the signal of PBH ALP is less sensitive to the given coupling constants of axion because its flux is independent of the couplings. By contrast, both flux and cross-section of solar axion are proportional to the coupling squared.

\subsection{The nucleon dipole portal}

Finally, we consider a dynamic interaction for axion inducing an analogous signal. The so-called axion-dependent nucleon electric dipole moment (EDM) is given by the following Lagrangian~\cite{Lucente:2022vuo}
\begin{equation}
\mathcal{L}_a^{\rm{nEDM}}=-\frac{i}{2}g_{d,N}a\bar{N}\gamma_5\sigma_{\mu\nu}NF^{\mu\nu}\;,
\end{equation}
where $N$ denotes proton or neutron. There is a model-independent feature for coupling $g_{d,N}$ similar to QCD axion, i.e., $g_d\equiv g_{d,n}=-g_{d,p}$. Consequently, we have a new detection channel, that is $a+p\to p+\gamma$ which implies that an axion can be absorbed by free proton and then emits a free photon. This reaction produces a visible flux of photons within detector. This hypothesis was used to predict the signal of axions emitted from supernova in free-proton-rich water Cherenkov detector such as Hyper-K~\cite{Lucente:2022vuo}. After integrating the phase space, the total cross-section of scattering process is given by the neat formula
\begin{eqnarray}
\sigma_{\rm{nEDM}}=g_d^2E_a^2/2\pi
\end{eqnarray}
under the limit of small axion mass, which does not depend on axion mass as shown in Fig.~\ref{fig:cro-section}. Considering PBH as the source of incoming axion, the total event rate will be only proportional to $g_d^2$. In the left panel of Fig.~\ref{fig:EventRate} (green line), we show the event rate of nEDM induced scattering for free proton.
The scattered target in this case is the hydrogen atom in water molecule in which we take into account two free protons.
One can see that the energy range sensitive to nEDM overlaps that of inverse Primakoff for $E_a>1$ MeV and peaks around several tens of MeV. Thus, this detection channel can also probe the PBH ALPs with energies around 100 MeV.

\subsection{Relevant constraints}
\label{sec:Constraint}

\subsubsection{The ALP lifetime}
\label{sec:decay}

Given the decay of ALP to diphoton, the lifetime of ALP becomes
\begin{eqnarray}
\tau_a={64\pi\over m_a^3 g_{a\gamma}^2}\;.
\end{eqnarray}
We require the ALP lifetime to be longer than the era of matter-radiation equality, i.e. $10^{11}~{\rm s}\sim 1~{\rm kpc}$, to give a conservative constraint on $g_{a\gamma}$.
Note that taking into account the relativistic Lorentz factor $\gamma\simeq E_a/m_a$ would make the constraint much weaker when $m_a\ll E_a$. Given $E_a\simeq 10$ MeV and $m_a=1$ keV for illustration, the limit on $g_{a\gamma}$ will be weakened by two orders of magnitude.

\subsubsection{The attenuation of ALPs}
\label{sec:attenuation}

The PBH ALPs can travel a distance in the atmosphere and underground before reaching the terrestrial detectors and would be stopped due to the process $a+p\to p+\gamma$ therein.
Their path $\ell$ is thus required to satisfy~\cite{Lucente:2022vuo}
\begin{eqnarray}
\ell=\Big(n_N \sigma_{\rm nEDM}\Big)^{-1} > d^\oplus ~ {\rm or} ~ d^{\rm atm}\;,
\end{eqnarray}
where $n_N$ is the nucleon density, $d^\oplus\simeq 1$ km is the depth of the detector from the Earth surface and $d^{\rm atm}=86$ km is the altitude of atmosphere in the US Standard Atmosphere model~\cite{Emken:2018run}. Taking a constant mass density of Earth crust $\rho^\oplus=2.7~{\rm g}~{\rm cm}^{-3}$~\cite{Emken:2018run} and running over the most abundant elements in the Earth, we obtain $n_N^\oplus=1.6\times 10^{24}~{\rm cm}^{-3}$. The atmosphere was modeled as 4 layers of falling mass density as a function of the altitude. We take a constant density $\rho^{\rm atm}=1.1~{\rm kg}~{\rm m}^{-3}$
in the first layer~\cite{Emken:2018run} and the most abundant elements in the atmosphere
(nitrogen and oxygen) to obtain the number density of nucleons as $n_N^{\rm atm}=6.5\times 10^{20}~{\rm cm}^{-3}$. Taking $E_a>100$ MeV as a reasonable spectrum limit, we find the upper limit of $g_d$ as $3~(17)~{\rm GeV}^{-2}$ from the attenuation underground (in the atmosphere) and thus the constraint due to the attenuation is very weak.

\subsubsection{The supernova bounds and others}
\label{sec:sn}

The supernova is also a novel source of ALPs and the relevant observations of supernova (SN) permit to place constraints on the ALP couplings.
There exists study of supernova population with low explosion energies placing the constraint on $a\to \gamma\gamma$ and the coupling $g_{a\gamma}$~\cite{Caputo:2022mah}.
The observation of the SN 1987A neutrino burst also allows to constrain the axion coupling $g_d$ by evaluating the axion luminosity~\cite{Lucente:2022vuo}. There are also other constraints on $g_{a\gamma}$ from visible decays of ALPs produced in
SN 1987A and beam dum experiments.
We place the relevant bounds in the figures below.

\section{Detection and constraints in neutrino experiments}
\label{sec:Detection}

Next we examine the detection of PBH ALP events and the relevant constraints in neutrino experiments.
The event number per recoil energy is given by
\begin{eqnarray}
{dN\over dE_R}=M_{\rm det}t {dR\over dE_R}\;,
\end{eqnarray}
where $M_{\rm det}$ is the fiducial mass of the detector and $t$ is the exposure time. As the neutrino experiments are only sensitive to visible energy above 100 MeV, we ignore the negligible AE absorption in $dR/dE_R$ and only consider the inverse Primakoff process for ALP coupling $g_{a\gamma}$ with water molecule or for nEDM hypothesis with free protons.

The Super-K detector performed 2628.1 days of running in a single tank consisting of 22.5 kiloton (kton) fiducial volume pure water. This corresponds to an exposure of 162 kiloton$\cdot$years. In the Super-K analysis, there are three energy bins with $100~{\rm MeV}<E<1.33~{\rm GeV}$, $1.33~{\rm GeV}<E<20~{\rm GeV}$ and $E>20~{\rm GeV}$ for the ``electron elastic scatter-like'' events~\cite{Super-Kamiokande:2017dch}. The numbers of measured event and the estimated background for each energy bin are listed in Table~I of Ref.~\cite{Super-Kamiokande:2017dch}.
To set bounds on the PBH ALP parameter space from Super-K, we define the $\chi^2$ with regard to the three energy bins
\begin{eqnarray}
\chi^2=\sum_{i=1}^3 \min_{\alpha_i} \left[2\left(N_\text{th}^i-N_\text{obs}^i+N_\text{obs}^i\ln\frac{N_\text{obs}^i}{N_\text{th}^i}\right)+\left(\frac{\alpha_i}{\sigma_i}\right)^2\right]\;,
\end{eqnarray}
where $N_\text{th}^i=N_\text{bkg}^i(1+\alpha_i)+N_\text{sig}^i$. Here $N_\text{bkg}^i$ ($N_\text{obs}^i$) [$N_\text{sig}^i$] is the number of background (observed) [signal] events in each bin, $\alpha_i$ is the background normalization factor that is minimized over, and $\sigma_i$ is the uncertainty for each bin. Follow Ref.~\cite{Super-Kamiokande:2017dch}, we take a conservative value of $19\%$ for the uncertainties in the two bins with lower energies, and 23\% for the uncertainty in the third energy bin. When calculating our signal event number, we also convolute the signal efficiency taken from the Fig.~1 in Ref.~\cite{Super-Kamiokande:2017dch}. As the event rate of inverse Primakoff decreases rapidly for energies above $\sim 100$ MeV due to the ALP flux from PBH, only the signal event for the first energy bin in fact matters for our analysis. The background-only hypothesis yields a $\chi^2_{0}=2.7$, and the 2$\sigma$ exclusion bound is set with $\Delta \chi^2=\chi^2-\chi^2_{0}=4.0$  for one degree of freedom~\cite{Li:2022jxo}.

For the Hyper-K with underground water Cherenkov detector, the fiducial mass
is 374 kton and we refer the exposure as 1870 kton$\cdot$years~\cite{Hyper-Kamiokande:2018ofw} with an ideal detection efficiency of signal as $\epsilon=1$. Following Ref.~\cite{Lucente:2022vuo}, we place a
conservative bound on PBH ALP from Hyper-K as
\begin{equation}
\epsilon\times N_s<N_{\rm HK}\;,
\label{equ:condition}
\end{equation}
where $N_s$ is the signal event number for $E>100$ MeV and $N_{\rm HK}=2$ is the assumed number of observation.

\begin{figure}[htb!]
\begin{center}
\includegraphics[scale=1,width=0.49\linewidth]{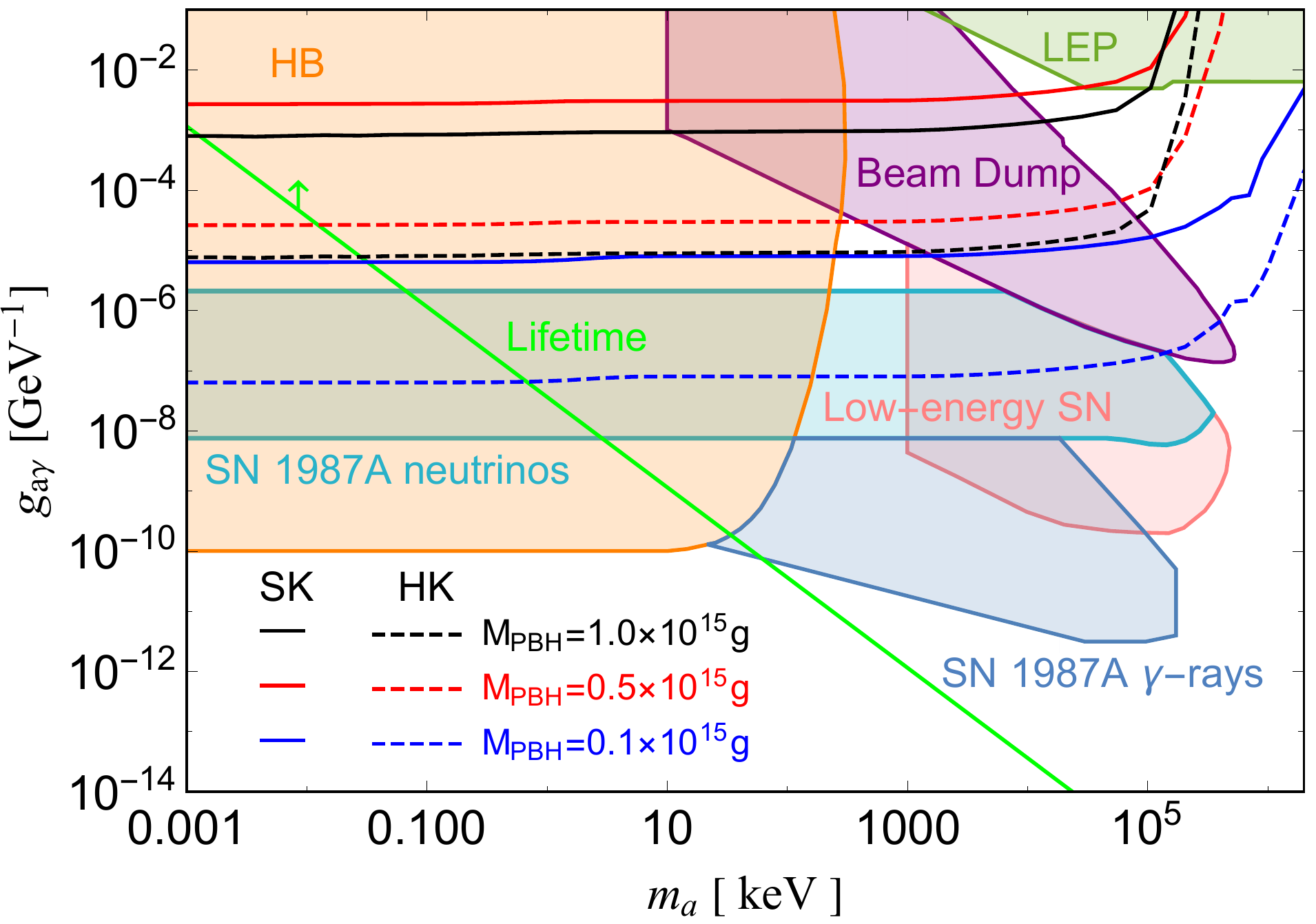}
\includegraphics[scale=1,width=0.49\linewidth]{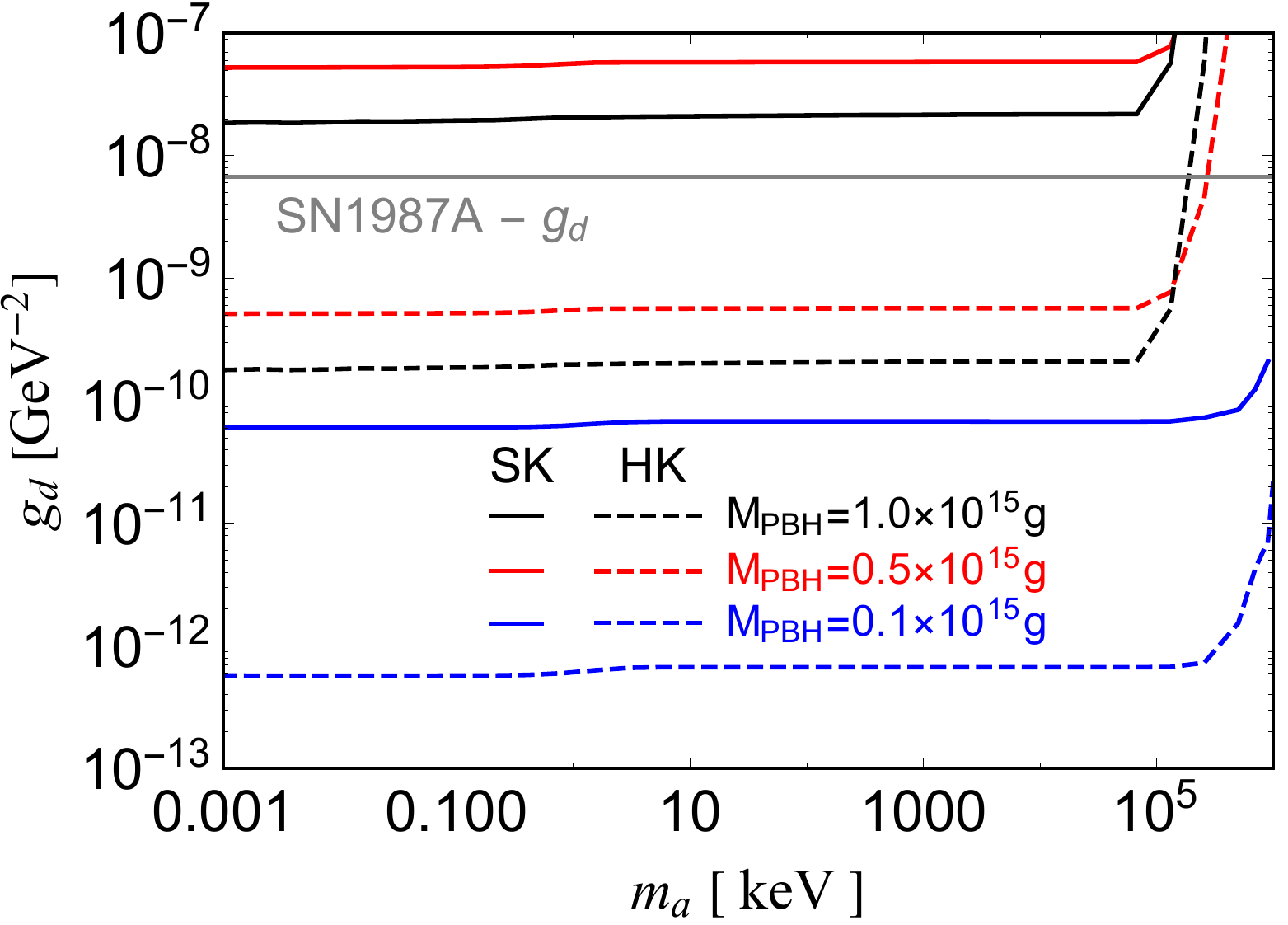}
\end{center}
\caption{The bounds from Super-K (solid) and projected sensitivity at Hyper-K (dashed) on $g_{a\gamma}$ (left) and $g_d$ (right) as a function of $m_a$. The PBH masses are assumed as $M_{\rm PBH}=1.0\times10^{15}$ g (black), $0.5\times10^{15}$ g (red) and $0.1\times10^{15}$ g (blue) with corresponding maximum $f_{\rm PBH}$ allowed by the extragalactic gamma-ray constraints. We also present other constraints on $g_{a\gamma}$ from low-energy supernovae radiation~\cite{Caputo:2022mah} (pink), beam dump experiments~\cite{Dolan_2017} (purple), SN 1987A gamma-rays~\cite{Caputo_2022,Jaeckel_2018} (sky blue), LEP (yellow green), horizontal branch stars~\cite{Jaeckel_2018} (orange), SN 1987A neutrinos~\cite{Caputo_2022,Caputo:2022mah} (teal) and ALP lifetime (green).
The upper limit of $g_d=6.7\times 10^{-9}~\rm{GeV}^{-2}$~\cite{Lucente:2022vuo} obtained by neutrino from SN1987A absorbed by proton is also shown in gray.
}
\label{fig:Bound}
\end{figure}

We first show the bounds on the ALP couplings versus $m_a$ from Super-K (solid line) and Hyper-K (dashed line) in Fig.~\ref{fig:Bound} (left panel for $g_{a\gamma}$ and right panel for $g_{d}$) for different choices of PBH masses with corresponding maximum $f_{\rm PBH}$ allowed by the extragalactic gamma-ray constraints.
It turns out that Super-K (Hyper-K) can confine the axion-photon coupling $g_{a\gamma}$ down to $8\times 10^{-4}~{\rm GeV}^{-1}$ ($8\times 10^{-6}~{\rm GeV}^{-1}$), $3\times 10^{-3}~{\rm GeV}^{-1}$ ($3\times 10^{-5}~{\rm GeV}^{-1}$) and $6\times 10^{-6}~{\rm GeV}^{-1}$ ($6\times 10^{-8}~{\rm GeV}^{-1}$) for $M_{\rm PBH}=1.0\times10^{15},~0.5\times10^{15},~0.1\times10^{15}$ g, respectively. We also display the constraints from ALP lifetime, supernova 1987A, the cooling of horizontal branch (HB) stars as well as beam dump experiments. It turns out that the above sensitive region of $g_{a\gamma}$ from PBH ALP has already been excluded by these constraints. For axion-nEDM coupling $g_d$, the corresponding limits become $2\times 10^{-8}~{\rm GeV}^{-2}$ ($2\times 10^{-10}~{\rm GeV}^{-2}$), $5\times 10^{-8}~{\rm GeV}^{-2}$ ($5\times 10^{-10}~{\rm GeV}^{-2}$) and $6\times 10^{-11}~{\rm GeV}^{-2}$ ($6\times 10^{-13}~{\rm GeV}^{-2}$) from Super-K (Hyper-K).

\begin{figure}[htb!]
\begin{center}
\includegraphics[scale=1,width=0.49\linewidth]{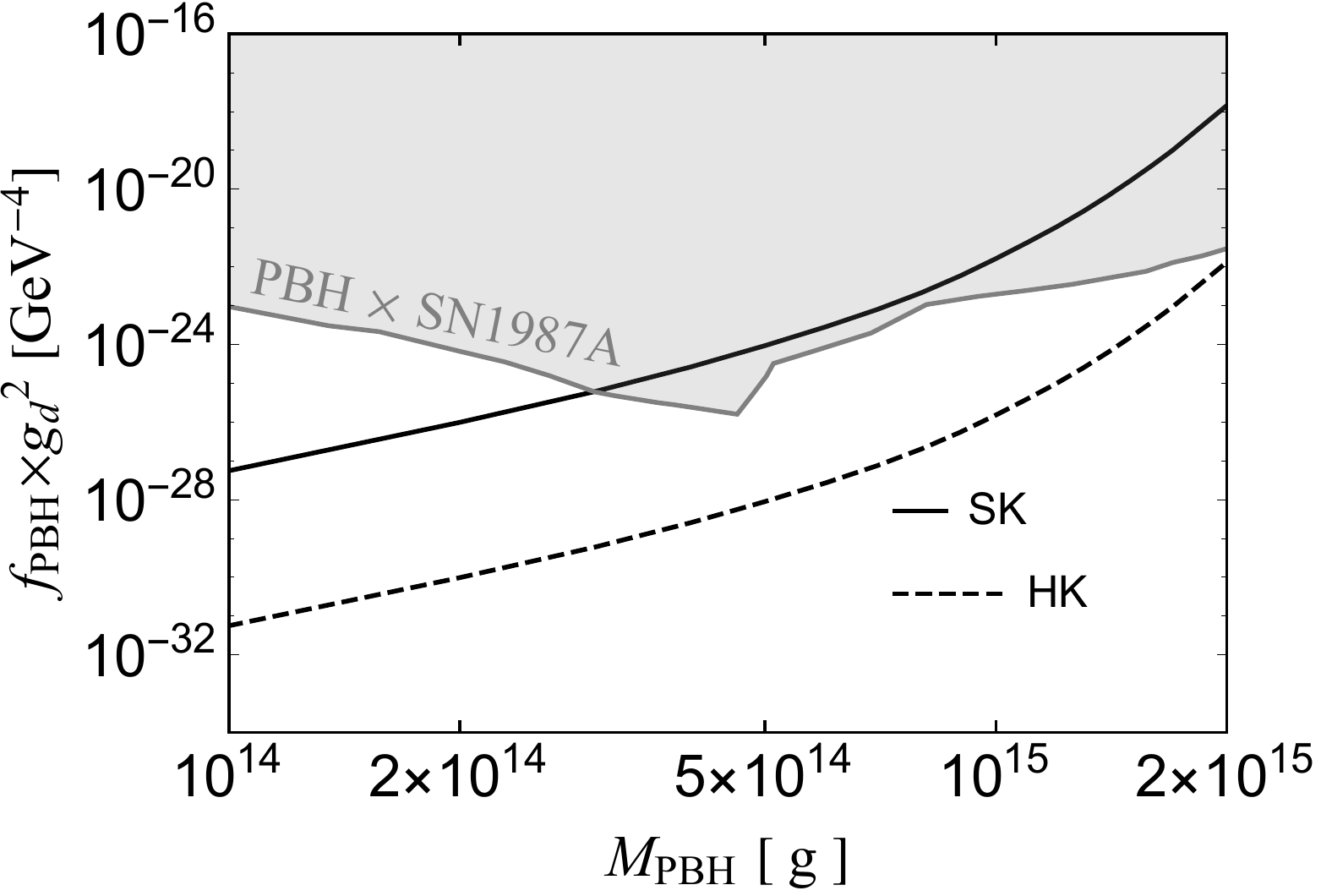}
\end{center}
\caption{The bounds on $f_{\rm PBH}\cdot g_d^2$ as a function of $M_{\rm PBH}$ from Super-K (solid) and Hyper-K (dashed) for $m_a=1$ keV as an illustration.
The current evaporation constraint on $f_{\rm PBH}$ from extragalactic gamma-rays~\cite{Carr:2020gox} and the SN 1987A bound on $g_d$~\cite{Lucente:2022vuo} are combined and the excluded region is shown in gray for comparison.}
\label{fig:PBHBound}
\end{figure}

These bounds can be translated into the upper limit on $f_{\rm PBH}\cdot g^2_{a\gamma(d)}$ as a function of $M_{\rm PBH}$, as shown in Fig.~\ref{fig:PBHBound} for $f_{\rm PBH}\cdot g^2_{d}$ and $m_a=1$ keV for illustration. The combined excluded region of $f_{\rm PBH}\cdot g^2_{d}$ from extragalactic gamma-rays for $f_{\rm PBH}$~\cite{Carr:2020gox} and SN 1987A for $g_{d}$~\cite{Lucente:2022vuo} is shown in gray.
One can see that the neutrino experiments can provide viable search for the axion-nEDM coupling of PBH ALP and are more sensitive to low PBH masses due to the visible energy range. As shown above, $g_{a\gamma}$ is highly constrained and we do not show the limit on $f_{\rm PBH}\cdot g^2_{a\gamma}$ here.

\section{Conclusion}
\label{sec:Con}

We explore the possibility that the light axion-like particles are produced from PBH evaporation. The ALPs receive energies in the region of a few hundred MeV. The absorption of light ALPs induces energetic photoelectron signatures in the underground detector targets of neutrino experiments. Given the galactic and extragalactic PBH ALP flux, we calculate the PBH ALP event rate produced by the inverse Primakoff process with axion-photon coupling $g_{a\gamma}$ and with axion-nucleon coupling $g_d$ for the hypothesis of axion-dependent nucleon electric dipole moment. We then place general exclusion limits on the axion couplings utilizing the Super-K data and predict the prospects for the future Hyper-K. We also translate these limits into the upper bound on the fraction of DM composed of PBHs $f_{\rm PBH}$.

We find the following conclusions.
\begin{itemize}
\item Super-K can constrain the axion-photon coupling $g_{a\gamma}$ as small as $8\times 10^{-4}~{\rm GeV}^{-1}$, $3\times 10^{-3}~{\rm GeV}^{-1}$ and $6\times 10^{-6}~{\rm GeV}^{-1}$ for $M_{\rm PBH}=1.0\times10^{15},~0.5\times10^{15},~0.1\times10^{15}$ g, respectively. The above sensitive region of $g_{a\gamma}$ from PBH ALP has already been excluded by other observations and constraints. For axion-nEDM coupling $g_d$, the corresponding limits become $2\times 10^{-8}~{\rm GeV}^{-2}$, $5\times 10^{-8}~{\rm GeV}^{-2}$ and $6\times 10^{-11}~{\rm GeV}^{-2}$ from Super-K.
\item Hyper-K can improve the limits of axion coupling constants by two orders of magnitude.
\item
The neutrino experiments provide improved bounds on $f_{\rm PBH}\cdot g_d^2$ for small PBH masses, compared with the constraints on $f_{\rm PBH}$ from extragalactic gamma-rays and $g_d$ from SN 1987A.
\end{itemize}
Note that these bounds on axion couplings and $f_{\rm PBH}$ are obtained simultaneously and are
fully correlated.


{\bf{Note Added:}}
During the completion of this work, a study~\cite{Cui:2022owf} appeared and investigated the boosted ALPs from dark matter decay in neutrino experiments.

\section*{ACKNOWLEDGMENTS}
We would like to thank Andrea Caputo and Wei Chao for useful communication.
T.L. is supported by the National Natural Science Foundation of China (Grant No. 11975129, 12035008) and ``the Fundamental Research Funds for the Central Universities'', Nankai University (Grants No. 63196013).

\bibliography{refs}

\begin{thebibliography}{82}%
\makeatletter
\providecommand \@ifxundefined [1]{%
 \@ifx{#1\undefined}
}%
\providecommand \@ifnum [1]{%
 \ifnum #1\expandafter \@firstoftwo
 \else \expandafter \@secondoftwo
 \fi
}%
\providecommand \@ifx [1]{%
 \ifx #1\expandafter \@firstoftwo
 \else \expandafter \@secondoftwo
 \fi
}%
\providecommand \natexlab [1]{#1}%
\providecommand \enquote  [1]{``#1''}%
\providecommand \bibnamefont  [1]{#1}%
\providecommand \bibfnamefont [1]{#1}%
\providecommand \citenamefont [1]{#1}%
\providecommand \href@noop [0]{\@secondoftwo}%
\providecommand \href [0]{\begingroup \@sanitize@url \@href}%
\providecommand \@href[1]{\@@startlink{#1}\@@href}%
\providecommand \@@href[1]{\endgroup#1\@@endlink}%
\providecommand \@sanitize@url [0]{\catcode `\\12\catcode `\$12\catcode
  `\&12\catcode `\#12\catcode `\^12\catcode `\_12\catcode `\%12\relax}%
\providecommand \@@startlink[1]{}%
\providecommand \@@endlink[0]{}%
\providecommand \url  [0]{\begingroup\@sanitize@url \@url }%
\providecommand \@url [1]{\endgroup\@href {#1}{\urlprefix }}%
\providecommand \urlprefix  [0]{URL }%
\providecommand \Eprint [0]{\href }%
\providecommand \doibase [0]{http://dx.doi.org/}%
\providecommand \selectlanguage [0]{\@gobble}%
\providecommand \bibinfo  [0]{\@secondoftwo}%
\providecommand \bibfield  [0]{\@secondoftwo}%
\providecommand \translation [1]{[#1]}%
\providecommand \BibitemOpen [0]{}%
\providecommand \bibitemStop [0]{}%
\providecommand \bibitemNoStop [0]{.\EOS\space}%
\providecommand \EOS [0]{\spacefactor3000\relax}%
\providecommand \BibitemShut  [1]{\csname bibitem#1\endcsname}%
\let\auto@bib@innerbib\@empty
\bibitem [{\citenamefont {Aalbers}\ \emph {et~al.}(2022)\citenamefont {Aalbers}
  \emph {et~al.}}]{LUX-ZEPLIN:2022qhg}%
  \BibitemOpen
  \bibfield  {author} {\bibinfo {author} {\bibfnamefont {J.}~\bibnamefont
  {Aalbers}} \emph {et~al.} (\bibinfo {collaboration} {LUX-ZEPLIN}),\
  }\href@noop {} {\  (\bibinfo {year} {2022})},\ \Eprint
  {http://arxiv.org/abs/2207.03764} {arXiv:2207.03764 [hep-ex]} \BibitemShut
  {NoStop}%
\bibitem [{\citenamefont {Aprile}\ \emph {et~al.}(2022)\citenamefont {Aprile}
  \emph {et~al.}}]{Aprile:2022vux}%
  \BibitemOpen
  \bibfield  {author} {\bibinfo {author} {\bibfnamefont {E.}~\bibnamefont
  {Aprile}} \emph {et~al.},\ }\href@noop {} {\  (\bibinfo {year} {2022})},\
  \Eprint {http://arxiv.org/abs/2207.11330} {arXiv:2207.11330 [hep-ex]}
  \BibitemShut {NoStop}%
\bibitem [{\citenamefont {Baluni}(1979)}]{Baluni:1978rf}%
  \BibitemOpen
  \bibfield  {author} {\bibinfo {author} {\bibfnamefont {V.}~\bibnamefont
  {Baluni}},\ }\href {\doibase 10.1103/PhysRevD.19.2227} {\bibfield  {journal}
  {\bibinfo  {journal} {Phys. Rev. D}\ }\textbf {\bibinfo {volume} {19}},\
  \bibinfo {pages} {2227} (\bibinfo {year} {1979})}\BibitemShut {NoStop}%
\bibitem [{\citenamefont {Crewther}\ \emph {et~al.}(1979)\citenamefont
  {Crewther}, \citenamefont {Di~Vecchia}, \citenamefont {Veneziano},\ and\
  \citenamefont {Witten}}]{Crewther:1979pi}%
  \BibitemOpen
  \bibfield  {author} {\bibinfo {author} {\bibfnamefont {R.~J.}\ \bibnamefont
  {Crewther}}, \bibinfo {author} {\bibfnamefont {P.}~\bibnamefont
  {Di~Vecchia}}, \bibinfo {author} {\bibfnamefont {G.}~\bibnamefont
  {Veneziano}}, \ and\ \bibinfo {author} {\bibfnamefont {E.}~\bibnamefont
  {Witten}},\ }\href {\doibase 10.1016/0370-2693(79)90128-X} {\bibfield
  {journal} {\bibinfo  {journal} {Phys. Lett. B}\ }\textbf {\bibinfo {volume}
  {88}},\ \bibinfo {pages} {123} (\bibinfo {year} {1979})},\ \bibinfo {note}
  {[Erratum: Phys.Lett.B 91, 487 (1980)]}\BibitemShut {NoStop}%
\bibitem [{\citenamefont {Kim}(1979)}]{Kim:1979if}%
  \BibitemOpen
  \bibfield  {author} {\bibinfo {author} {\bibfnamefont {J.~E.}\ \bibnamefont
  {Kim}},\ }\href {\doibase 10.1103/PhysRevLett.43.103} {\bibfield  {journal}
  {\bibinfo  {journal} {Phys. Rev. Lett.}\ }\textbf {\bibinfo {volume} {43}},\
  \bibinfo {pages} {103} (\bibinfo {year} {1979})}\BibitemShut {NoStop}%
\bibitem [{\citenamefont {Shifman}\ \emph {et~al.}(1980)\citenamefont
  {Shifman}, \citenamefont {Vainshtein},\ and\ \citenamefont
  {Zakharov}}]{Shifman:1979if}%
  \BibitemOpen
  \bibfield  {author} {\bibinfo {author} {\bibfnamefont {M.~A.}\ \bibnamefont
  {Shifman}}, \bibinfo {author} {\bibfnamefont {A.~I.}\ \bibnamefont
  {Vainshtein}}, \ and\ \bibinfo {author} {\bibfnamefont {V.~I.}\ \bibnamefont
  {Zakharov}},\ }\href {\doibase 10.1016/0550-3213(80)90209-6} {\bibfield
  {journal} {\bibinfo  {journal} {Nucl. Phys. B}\ }\textbf {\bibinfo {volume}
  {166}},\ \bibinfo {pages} {493} (\bibinfo {year} {1980})}\BibitemShut
  {NoStop}%
\bibitem [{\citenamefont {Dine}\ \emph {et~al.}(1981)\citenamefont {Dine},
  \citenamefont {Fischler},\ and\ \citenamefont {Srednicki}}]{Dine:1981rt}%
  \BibitemOpen
  \bibfield  {author} {\bibinfo {author} {\bibfnamefont {M.}~\bibnamefont
  {Dine}}, \bibinfo {author} {\bibfnamefont {W.}~\bibnamefont {Fischler}}, \
  and\ \bibinfo {author} {\bibfnamefont {M.}~\bibnamefont {Srednicki}},\ }\href
  {\doibase 10.1016/0370-2693(81)90590-6} {\bibfield  {journal} {\bibinfo
  {journal} {Phys. Lett. B}\ }\textbf {\bibinfo {volume} {104}},\ \bibinfo
  {pages} {199} (\bibinfo {year} {1981})}\BibitemShut {NoStop}%
\bibitem [{\citenamefont {Zhitnitsky}(1980)}]{Zhitnitsky:1980tq}%
  \BibitemOpen
  \bibfield  {author} {\bibinfo {author} {\bibfnamefont {A.~R.}\ \bibnamefont
  {Zhitnitsky}},\ }\href@noop {} {\bibfield  {journal} {\bibinfo  {journal}
  {Sov. J. Nucl. Phys.}\ }\textbf {\bibinfo {volume} {31}},\ \bibinfo {pages}
  {260} (\bibinfo {year} {1980})}\BibitemShut {NoStop}%
\bibitem [{\citenamefont {Baker}\ \emph {et~al.}(2006)\citenamefont {Baker}
  \emph {et~al.}}]{Baker:2006ts}%
  \BibitemOpen
  \bibfield  {author} {\bibinfo {author} {\bibfnamefont {C.~A.}\ \bibnamefont
  {Baker}} \emph {et~al.},\ }\href {\doibase 10.1103/PhysRevLett.97.131801}
  {\bibfield  {journal} {\bibinfo  {journal} {Phys. Rev. Lett.}\ }\textbf
  {\bibinfo {volume} {97}},\ \bibinfo {pages} {131801} (\bibinfo {year}
  {2006})},\ \Eprint {http://arxiv.org/abs/hep-ex/0602020}
  {arXiv:hep-ex/0602020} \BibitemShut {NoStop}%
\bibitem [{\citenamefont {Pendlebury}\ \emph {et~al.}(2015)\citenamefont
  {Pendlebury} \emph {et~al.}}]{Pendlebury:2015lrz}%
  \BibitemOpen
  \bibfield  {author} {\bibinfo {author} {\bibfnamefont {J.~M.}\ \bibnamefont
  {Pendlebury}} \emph {et~al.},\ }\href {\doibase 10.1103/PhysRevD.92.092003}
  {\bibfield  {journal} {\bibinfo  {journal} {Phys. Rev. D}\ }\textbf {\bibinfo
  {volume} {92}},\ \bibinfo {pages} {092003} (\bibinfo {year} {2015})},\
  \Eprint {http://arxiv.org/abs/1509.04411} {arXiv:1509.04411 [hep-ex]}
  \BibitemShut {NoStop}%
\bibitem [{\citenamefont {Peccei}\ and\ \citenamefont
  {Quinn}(1977{\natexlab{a}})}]{Peccei:1977hh}%
  \BibitemOpen
  \bibfield  {author} {\bibinfo {author} {\bibfnamefont {R.~D.}\ \bibnamefont
  {Peccei}}\ and\ \bibinfo {author} {\bibfnamefont {H.~R.}\ \bibnamefont
  {Quinn}},\ }\href {\doibase 10.1103/PhysRevLett.38.1440} {\bibfield
  {journal} {\bibinfo  {journal} {Phys. Rev. Lett.}\ }\textbf {\bibinfo
  {volume} {38}},\ \bibinfo {pages} {1440} (\bibinfo {year}
  {1977}{\natexlab{a}})}\BibitemShut {NoStop}%
\bibitem [{\citenamefont {Peccei}\ and\ \citenamefont
  {Quinn}(1977{\natexlab{b}})}]{Peccei:1977ur}%
  \BibitemOpen
  \bibfield  {author} {\bibinfo {author} {\bibfnamefont {R.~D.}\ \bibnamefont
  {Peccei}}\ and\ \bibinfo {author} {\bibfnamefont {H.~R.}\ \bibnamefont
  {Quinn}},\ }\href {\doibase 10.1103/PhysRevD.16.1791} {\bibfield  {journal}
  {\bibinfo  {journal} {Phys. Rev. D}\ }\textbf {\bibinfo {volume} {16}},\
  \bibinfo {pages} {1791} (\bibinfo {year} {1977}{\natexlab{b}})}\BibitemShut
  {NoStop}%
\bibitem [{\citenamefont {Weinberg}(1978)}]{Weinberg:1977ma}%
  \BibitemOpen
  \bibfield  {author} {\bibinfo {author} {\bibfnamefont {S.}~\bibnamefont
  {Weinberg}},\ }\href {\doibase 10.1103/PhysRevLett.40.223} {\bibfield
  {journal} {\bibinfo  {journal} {Phys. Rev. Lett.}\ }\textbf {\bibinfo
  {volume} {40}},\ \bibinfo {pages} {223} (\bibinfo {year} {1978})}\BibitemShut
  {NoStop}%
\bibitem [{\citenamefont {Wilczek}(1978)}]{Wilczek:1977pj}%
  \BibitemOpen
  \bibfield  {author} {\bibinfo {author} {\bibfnamefont {F.}~\bibnamefont
  {Wilczek}},\ }\href {\doibase 10.1103/PhysRevLett.40.279} {\bibfield
  {journal} {\bibinfo  {journal} {Phys. Rev. Lett.}\ }\textbf {\bibinfo
  {volume} {40}},\ \bibinfo {pages} {279} (\bibinfo {year} {1978})}\BibitemShut
  {NoStop}%
\bibitem [{\citenamefont {Di~Luzio}\ \emph
  {et~al.}(2020{\natexlab{a}})\citenamefont {Di~Luzio}, \citenamefont
  {Giannotti}, \citenamefont {Nardi},\ and\ \citenamefont
  {Visinelli}}]{DiLuzio:2020wdo}%
  \BibitemOpen
  \bibfield  {author} {\bibinfo {author} {\bibfnamefont {L.}~\bibnamefont
  {Di~Luzio}}, \bibinfo {author} {\bibfnamefont {M.}~\bibnamefont {Giannotti}},
  \bibinfo {author} {\bibfnamefont {E.}~\bibnamefont {Nardi}}, \ and\ \bibinfo
  {author} {\bibfnamefont {L.}~\bibnamefont {Visinelli}},\ }\href {\doibase
  10.1016/j.physrep.2020.06.002} {\bibfield  {journal} {\bibinfo  {journal}
  {Phys. Rept.}\ }\textbf {\bibinfo {volume} {870}},\ \bibinfo {pages} {1}
  (\bibinfo {year} {2020}{\natexlab{a}})},\ \Eprint
  {http://arxiv.org/abs/2003.01100} {arXiv:2003.01100 [hep-ph]} \BibitemShut
  {NoStop}%
\bibitem [{\citenamefont {Kim}(1987)}]{Kim:1986ax}%
  \BibitemOpen
  \bibfield  {author} {\bibinfo {author} {\bibfnamefont {J.~E.}\ \bibnamefont
  {Kim}},\ }\href {\doibase 10.1016/0370-1573(87)90017-2} {\bibfield  {journal}
  {\bibinfo  {journal} {Phys. Rept.}\ }\textbf {\bibinfo {volume} {150}},\
  \bibinfo {pages} {1} (\bibinfo {year} {1987})}\BibitemShut {NoStop}%
\bibitem [{\citenamefont {Kuster}\ \emph {et~al.}(2008)\citenamefont {Kuster},
  \citenamefont {Raffelt},\ and\ \citenamefont {Beltran}}]{Kuster:2008zz}%
  \BibitemOpen
  \bibinfo {editor} {\bibfnamefont {M.}~\bibnamefont {Kuster}}, \bibinfo
  {editor} {\bibfnamefont {G.}~\bibnamefont {Raffelt}}, \ and\ \bibinfo
  {editor} {\bibfnamefont {B.}~\bibnamefont {Beltran}},\ eds.,\ \href@noop {}
  {\emph {\bibinfo {title} {{Axions: Theory, cosmology, and experimental
  searches. Proceedings, 1st Joint ILIAS-CERN-CAST axion training, Geneva,
  Switzerland, November 30-December 2, 2005}}}},\ Vol.\ \bibinfo {volume}
  {741}\ (\bibinfo {year} {2008})\BibitemShut {NoStop}%
\bibitem [{\citenamefont {Zel'dovich}(1967)}]{Zeldovich:1967lct}%
  \BibitemOpen
  \bibfield  {author} {\bibinfo {author} {\bibfnamefont {Y.~B. .~N.}\
  \bibnamefont {Zel'dovich}, \bibfnamefont {I.~D.}},\ }\href@noop {} {\bibfield
   {journal} {\bibinfo  {journal} {Soviet Astron. AJ (Engl. Transl. ),}\
  }\textbf {\bibinfo {volume} {10}},\ \bibinfo {pages} {602} (\bibinfo {year}
  {1967})}\BibitemShut {NoStop}%
\bibitem [{\citenamefont {Carr}\ and\ \citenamefont
  {Hawking}(1974)}]{Carr:1974nx}%
  \BibitemOpen
  \bibfield  {author} {\bibinfo {author} {\bibfnamefont {B.~J.}\ \bibnamefont
  {Carr}}\ and\ \bibinfo {author} {\bibfnamefont {S.~W.}\ \bibnamefont
  {Hawking}},\ }\href@noop {} {\bibfield  {journal} {\bibinfo  {journal} {Mon.
  Not. Roy. Astron. Soc.}\ }\textbf {\bibinfo {volume} {168}},\ \bibinfo
  {pages} {399} (\bibinfo {year} {1974})}\BibitemShut {NoStop}%
\bibitem [{\citenamefont {Carr}(1975)}]{Carr:1975qj}%
  \BibitemOpen
  \bibfield  {author} {\bibinfo {author} {\bibfnamefont {B.~J.}\ \bibnamefont
  {Carr}},\ }\href {\doibase 10.1086/153853} {\bibfield  {journal} {\bibinfo
  {journal} {Astrophys. J.}\ }\textbf {\bibinfo {volume} {201}},\ \bibinfo
  {pages} {1} (\bibinfo {year} {1975})}\BibitemShut {NoStop}%
\bibitem [{\citenamefont {Khlopov}(2010)}]{Khlopov:2008qy}%
  \BibitemOpen
  \bibfield  {author} {\bibinfo {author} {\bibfnamefont {M.~Y.}\ \bibnamefont
  {Khlopov}},\ }\href {\doibase 10.1088/1674-4527/10/6/001} {\bibfield
  {journal} {\bibinfo  {journal} {Res. Astron. Astrophys.}\ }\textbf {\bibinfo
  {volume} {10}},\ \bibinfo {pages} {495} (\bibinfo {year} {2010})},\ \Eprint
  {http://arxiv.org/abs/0801.0116} {arXiv:0801.0116 [astro-ph]} \BibitemShut
  {NoStop}%
\bibitem [{\citenamefont {Carr}\ and\ \citenamefont
  {Kuhnel}(2021)}]{Carr:2021bzv}%
  \BibitemOpen
  \bibfield  {author} {\bibinfo {author} {\bibfnamefont {B.}~\bibnamefont
  {Carr}}\ and\ \bibinfo {author} {\bibfnamefont {F.}~\bibnamefont {Kuhnel}},\
  }in\ \href@noop {} {\emph {\bibinfo {booktitle} {{Les Houches summer school
  on Dark Matter}}}}\ (\bibinfo {year} {2021})\ \Eprint
  {http://arxiv.org/abs/2110.02821} {arXiv:2110.02821 [astro-ph.CO]}
  \BibitemShut {NoStop}%
\bibitem [{\citenamefont {Hawking}(1974)}]{Hawking:1974rv}%
  \BibitemOpen
  \bibfield  {author} {\bibinfo {author} {\bibfnamefont {S.~W.}\ \bibnamefont
  {Hawking}},\ }\href {\doibase 10.1038/248030a0} {\bibfield  {journal}
  {\bibinfo  {journal} {Nature}\ }\textbf {\bibinfo {volume} {248}},\ \bibinfo
  {pages} {30} (\bibinfo {year} {1974})}\BibitemShut {NoStop}%
\bibitem [{\citenamefont {Barrau}\ \emph {et~al.}(2004)\citenamefont {Barrau},
  \citenamefont {Blais}, \citenamefont {Boudoul},\ and\ \citenamefont
  {Polarski}}]{Barrau:2003xp}%
  \BibitemOpen
  \bibfield  {author} {\bibinfo {author} {\bibfnamefont {A.}~\bibnamefont
  {Barrau}}, \bibinfo {author} {\bibfnamefont {D.}~\bibnamefont {Blais}},
  \bibinfo {author} {\bibfnamefont {G.}~\bibnamefont {Boudoul}}, \ and\
  \bibinfo {author} {\bibfnamefont {D.}~\bibnamefont {Polarski}},\ }\href
  {\doibase 10.1002/andp.200310067} {\bibfield  {journal} {\bibinfo  {journal}
  {Annalen Phys.}\ }\textbf {\bibinfo {volume} {13}},\ \bibinfo {pages} {115}
  (\bibinfo {year} {2004})},\ \Eprint {http://arxiv.org/abs/astro-ph/0303330}
  {arXiv:astro-ph/0303330} \BibitemShut {NoStop}%
\bibitem [{\citenamefont {Carr}\ \emph {et~al.}(2020)\citenamefont {Carr},
  \citenamefont {Kohri}, \citenamefont {Sendouda},\ and\ \citenamefont
  {Yokoyama}}]{Carr:2020gox}%
  \BibitemOpen
  \bibfield  {author} {\bibinfo {author} {\bibfnamefont {B.}~\bibnamefont
  {Carr}}, \bibinfo {author} {\bibfnamefont {K.}~\bibnamefont {Kohri}},
  \bibinfo {author} {\bibfnamefont {Y.}~\bibnamefont {Sendouda}}, \ and\
  \bibinfo {author} {\bibfnamefont {J.}~\bibnamefont {Yokoyama}},\ }\href@noop
  {} {\  (\bibinfo {year} {2020})},\ \Eprint {http://arxiv.org/abs/2002.12778}
  {arXiv:2002.12778 [astro-ph.CO]} \BibitemShut {NoStop}%
\bibitem [{\citenamefont {Laha}(2019)}]{Laha:2019ssq}%
  \BibitemOpen
  \bibfield  {author} {\bibinfo {author} {\bibfnamefont {R.}~\bibnamefont
  {Laha}},\ }\href {\doibase 10.1103/PhysRevLett.123.251101} {\bibfield
  {journal} {\bibinfo  {journal} {Phys. Rev. Lett.}\ }\textbf {\bibinfo
  {volume} {123}},\ \bibinfo {pages} {251101} (\bibinfo {year} {2019})},\
  \Eprint {http://arxiv.org/abs/1906.09994} {arXiv:1906.09994 [astro-ph.HE]}
  \BibitemShut {NoStop}%
\bibitem [{\citenamefont {Laha}\ \emph {et~al.}(2020)\citenamefont {Laha},
  \citenamefont {Mu\~noz},\ and\ \citenamefont {Slatyer}}]{Laha:2020ivk}%
  \BibitemOpen
  \bibfield  {author} {\bibinfo {author} {\bibfnamefont {R.}~\bibnamefont
  {Laha}}, \bibinfo {author} {\bibfnamefont {J.~B.}\ \bibnamefont {Mu\~noz}}, \
  and\ \bibinfo {author} {\bibfnamefont {T.~R.}\ \bibnamefont {Slatyer}},\
  }\href {\doibase 10.1103/PhysRevD.101.123514} {\bibfield  {journal} {\bibinfo
   {journal} {Phys. Rev. D}\ }\textbf {\bibinfo {volume} {101}},\ \bibinfo
  {pages} {123514} (\bibinfo {year} {2020})},\ \Eprint
  {http://arxiv.org/abs/2004.00627} {arXiv:2004.00627 [astro-ph.CO]}
  \BibitemShut {NoStop}%
\bibitem [{\citenamefont {Saha}\ and\ \citenamefont
  {Laha}(2022)}]{Saha:2021pqf}%
  \BibitemOpen
  \bibfield  {author} {\bibinfo {author} {\bibfnamefont {A.~K.}\ \bibnamefont
  {Saha}}\ and\ \bibinfo {author} {\bibfnamefont {R.}~\bibnamefont {Laha}},\
  }\href {\doibase 10.1103/PhysRevD.105.103026} {\bibfield  {journal} {\bibinfo
   {journal} {Phys. Rev. D}\ }\textbf {\bibinfo {volume} {105}},\ \bibinfo
  {pages} {103026} (\bibinfo {year} {2022})},\ \Eprint
  {http://arxiv.org/abs/2112.10794} {arXiv:2112.10794 [astro-ph.CO]}
  \BibitemShut {NoStop}%
\bibitem [{\citenamefont {Ray}\ \emph {et~al.}(2021)\citenamefont {Ray},
  \citenamefont {Laha}, \citenamefont {Mu\~noz},\ and\ \citenamefont
  {Caputo}}]{Ray:2021mxu}%
  \BibitemOpen
  \bibfield  {author} {\bibinfo {author} {\bibfnamefont {A.}~\bibnamefont
  {Ray}}, \bibinfo {author} {\bibfnamefont {R.}~\bibnamefont {Laha}}, \bibinfo
  {author} {\bibfnamefont {J.~B.}\ \bibnamefont {Mu\~noz}}, \ and\ \bibinfo
  {author} {\bibfnamefont {R.}~\bibnamefont {Caputo}},\ }\href {\doibase
  10.1103/PhysRevD.104.023516} {\bibfield  {journal} {\bibinfo  {journal}
  {Phys. Rev. D}\ }\textbf {\bibinfo {volume} {104}},\ \bibinfo {pages}
  {023516} (\bibinfo {year} {2021})},\ \Eprint
  {http://arxiv.org/abs/2102.06714} {arXiv:2102.06714 [astro-ph.CO]}
  \BibitemShut {NoStop}%
\bibitem [{\citenamefont {Dasgupta}\ \emph {et~al.}(2020)\citenamefont
  {Dasgupta}, \citenamefont {Laha},\ and\ \citenamefont
  {Ray}}]{Dasgupta:2019cae}%
  \BibitemOpen
  \bibfield  {author} {\bibinfo {author} {\bibfnamefont {B.}~\bibnamefont
  {Dasgupta}}, \bibinfo {author} {\bibfnamefont {R.}~\bibnamefont {Laha}}, \
  and\ \bibinfo {author} {\bibfnamefont {A.}~\bibnamefont {Ray}},\ }\href
  {\doibase 10.1103/PhysRevLett.125.101101} {\bibfield  {journal} {\bibinfo
  {journal} {Phys. Rev. Lett.}\ }\textbf {\bibinfo {volume} {125}},\ \bibinfo
  {pages} {101101} (\bibinfo {year} {2020})},\ \Eprint
  {http://arxiv.org/abs/1912.01014} {arXiv:1912.01014 [hep-ph]} \BibitemShut
  {NoStop}%
\bibitem [{\citenamefont {Wang}\ \emph {et~al.}(2021)\citenamefont {Wang},
  \citenamefont {Xia}, \citenamefont {Zhang}, \citenamefont {Zhou},\ and\
  \citenamefont {Chang}}]{Wang:2020uvi}%
  \BibitemOpen
  \bibfield  {author} {\bibinfo {author} {\bibfnamefont {S.}~\bibnamefont
  {Wang}}, \bibinfo {author} {\bibfnamefont {D.-M.}\ \bibnamefont {Xia}},
  \bibinfo {author} {\bibfnamefont {X.}~\bibnamefont {Zhang}}, \bibinfo
  {author} {\bibfnamefont {S.}~\bibnamefont {Zhou}}, \ and\ \bibinfo {author}
  {\bibfnamefont {Z.}~\bibnamefont {Chang}},\ }\href {\doibase
  10.1103/PhysRevD.103.043010} {\bibfield  {journal} {\bibinfo  {journal}
  {Phys. Rev. D}\ }\textbf {\bibinfo {volume} {103}},\ \bibinfo {pages}
  {043010} (\bibinfo {year} {2021})},\ \Eprint
  {http://arxiv.org/abs/2010.16053} {arXiv:2010.16053 [hep-ph]} \BibitemShut
  {NoStop}%
\bibitem [{\citenamefont {Calabrese}\ \emph {et~al.}(2021)\citenamefont
  {Calabrese}, \citenamefont {Fiorillo}, \citenamefont {Miele}, \citenamefont
  {Morisi},\ and\ \citenamefont {Palazzo}}]{Calabrese:2021zfq}%
  \BibitemOpen
  \bibfield  {author} {\bibinfo {author} {\bibfnamefont {R.}~\bibnamefont
  {Calabrese}}, \bibinfo {author} {\bibfnamefont {D.~F.~G.}\ \bibnamefont
  {Fiorillo}}, \bibinfo {author} {\bibfnamefont {G.}~\bibnamefont {Miele}},
  \bibinfo {author} {\bibfnamefont {S.}~\bibnamefont {Morisi}}, \ and\ \bibinfo
  {author} {\bibfnamefont {A.}~\bibnamefont {Palazzo}},\ }\href@noop {} {\
  (\bibinfo {year} {2021})},\ \Eprint {http://arxiv.org/abs/2106.02492}
  {arXiv:2106.02492 [hep-ph]} \BibitemShut {NoStop}%
\bibitem [{\citenamefont {De~Romeri}\ \emph {et~al.}(2021)\citenamefont
  {De~Romeri}, \citenamefont {Mart\'\i{}nez-Mirav\'e},\ and\ \citenamefont
  {T\'ortola}}]{DeRomeri:2021xgy}%
  \BibitemOpen
  \bibfield  {author} {\bibinfo {author} {\bibfnamefont {V.}~\bibnamefont
  {De~Romeri}}, \bibinfo {author} {\bibfnamefont {P.}~\bibnamefont
  {Mart\'\i{}nez-Mirav\'e}}, \ and\ \bibinfo {author} {\bibfnamefont
  {M.}~\bibnamefont {T\'ortola}},\ }\href@noop {} {\  (\bibinfo {year}
  {2021})},\ \Eprint {http://arxiv.org/abs/2106.05013} {arXiv:2106.05013
  [hep-ph]} \BibitemShut {NoStop}%
\bibitem [{\citenamefont {Ghosh}\ \emph {et~al.}(2021)\citenamefont {Ghosh},
  \citenamefont {Guha},\ and\ \citenamefont {Sachdeva}}]{Ghosh:2021vkt}%
  \BibitemOpen
  \bibfield  {author} {\bibinfo {author} {\bibfnamefont {D.}~\bibnamefont
  {Ghosh}}, \bibinfo {author} {\bibfnamefont {A.}~\bibnamefont {Guha}}, \ and\
  \bibinfo {author} {\bibfnamefont {D.}~\bibnamefont {Sachdeva}},\ }\href@noop
  {} {\  (\bibinfo {year} {2021})},\ \Eprint {http://arxiv.org/abs/2110.00025}
  {arXiv:2110.00025 [hep-ph]} \BibitemShut {NoStop}%
\bibitem [{\citenamefont {Capanema}\ \emph {et~al.}(2021)\citenamefont
  {Capanema}, \citenamefont {Esmaeili},\ and\ \citenamefont
  {Esmaili}}]{Capanema:2021hnm}%
  \BibitemOpen
  \bibfield  {author} {\bibinfo {author} {\bibfnamefont {A.}~\bibnamefont
  {Capanema}}, \bibinfo {author} {\bibfnamefont {A.}~\bibnamefont {Esmaeili}},
  \ and\ \bibinfo {author} {\bibfnamefont {A.}~\bibnamefont {Esmaili}},\
  }\href@noop {} {\  (\bibinfo {year} {2021})},\ \Eprint
  {http://arxiv.org/abs/2110.05637} {arXiv:2110.05637 [hep-ph]} \BibitemShut
  {NoStop}%
\bibitem [{\citenamefont {Chao}\ \emph {et~al.}(2021)\citenamefont {Chao},
  \citenamefont {Li},\ and\ \citenamefont {Liao}}]{Chao:2021orr}%
  \BibitemOpen
  \bibfield  {author} {\bibinfo {author} {\bibfnamefont {W.}~\bibnamefont
  {Chao}}, \bibinfo {author} {\bibfnamefont {T.}~\bibnamefont {Li}}, \ and\
  \bibinfo {author} {\bibfnamefont {J.}~\bibnamefont {Liao}},\ }\href@noop {}
  {\  (\bibinfo {year} {2021})},\ \Eprint {http://arxiv.org/abs/2108.05608}
  {arXiv:2108.05608 [hep-ph]} \BibitemShut {NoStop}%
\bibitem [{\citenamefont {Bernal}\ \emph {et~al.}(2022)\citenamefont {Bernal},
  \citenamefont {Mu\~noz Albornoz}, \citenamefont {Palomares-Ruiz},\ and\
  \citenamefont {Villanueva-Domingo}}]{Bernal:2022swt}%
  \BibitemOpen
  \bibfield  {author} {\bibinfo {author} {\bibfnamefont {N.}~\bibnamefont
  {Bernal}}, \bibinfo {author} {\bibfnamefont {V.}~\bibnamefont {Mu\~noz
  Albornoz}}, \bibinfo {author} {\bibfnamefont {S.}~\bibnamefont
  {Palomares-Ruiz}}, \ and\ \bibinfo {author} {\bibfnamefont {P.}~\bibnamefont
  {Villanueva-Domingo}},\ }\href@noop {} {\  (\bibinfo {year} {2022})},\
  \Eprint {http://arxiv.org/abs/2203.14979} {arXiv:2203.14979 [hep-ph]}
  \BibitemShut {NoStop}%
\bibitem [{\citenamefont {Calabrese}\ \emph
  {et~al.}(2022{\natexlab{a}})\citenamefont {Calabrese}, \citenamefont
  {Chianese}, \citenamefont {Fiorillo},\ and\ \citenamefont
  {Saviano}}]{Calabrese:2021src}%
  \BibitemOpen
  \bibfield  {author} {\bibinfo {author} {\bibfnamefont {R.}~\bibnamefont
  {Calabrese}}, \bibinfo {author} {\bibfnamefont {M.}~\bibnamefont {Chianese}},
  \bibinfo {author} {\bibfnamefont {D.~F.~G.}\ \bibnamefont {Fiorillo}}, \ and\
  \bibinfo {author} {\bibfnamefont {N.}~\bibnamefont {Saviano}},\ }\href
  {\doibase 10.1103/PhysRevD.105.L021302} {\bibfield  {journal} {\bibinfo
  {journal} {Phys. Rev. D}\ }\textbf {\bibinfo {volume} {105}},\ \bibinfo
  {pages} {L021302} (\bibinfo {year} {2022}{\natexlab{a}})},\ \Eprint
  {http://arxiv.org/abs/2107.13001} {arXiv:2107.13001 [hep-ph]} \BibitemShut
  {NoStop}%
\bibitem [{\citenamefont {Li}\ and\ \citenamefont {Liao}(2022)}]{Li:2022jxo}%
  \BibitemOpen
  \bibfield  {author} {\bibinfo {author} {\bibfnamefont {T.}~\bibnamefont
  {Li}}\ and\ \bibinfo {author} {\bibfnamefont {J.}~\bibnamefont {Liao}},\
  }\href {\doibase 10.1103/PhysRevD.106.055043} {\bibfield  {journal} {\bibinfo
   {journal} {Phys. Rev. D}\ }\textbf {\bibinfo {volume} {106}},\ \bibinfo
  {pages} {055043} (\bibinfo {year} {2022})},\ \Eprint
  {http://arxiv.org/abs/2203.14443} {arXiv:2203.14443 [hep-ph]} \BibitemShut
  {NoStop}%
\bibitem [{\citenamefont {Calabrese}\ \emph
  {et~al.}(2022{\natexlab{b}})\citenamefont {Calabrese}, \citenamefont
  {Chianese}, \citenamefont {Fiorillo},\ and\ \citenamefont
  {Saviano}}]{Calabrese:2022rfa}%
  \BibitemOpen
  \bibfield  {author} {\bibinfo {author} {\bibfnamefont {R.}~\bibnamefont
  {Calabrese}}, \bibinfo {author} {\bibfnamefont {M.}~\bibnamefont {Chianese}},
  \bibinfo {author} {\bibfnamefont {D.~F.~G.}\ \bibnamefont {Fiorillo}}, \ and\
  \bibinfo {author} {\bibfnamefont {N.}~\bibnamefont {Saviano}},\ }\href
  {\doibase 10.1103/PhysRevD.105.103024} {\bibfield  {journal} {\bibinfo
  {journal} {Phys. Rev. D}\ }\textbf {\bibinfo {volume} {105}},\ \bibinfo
  {pages} {103024} (\bibinfo {year} {2022}{\natexlab{b}})},\ \Eprint
  {http://arxiv.org/abs/2203.17093} {arXiv:2203.17093 [hep-ph]} \BibitemShut
  {NoStop}%
\bibitem [{\citenamefont {Calz\`a}\ \emph {et~al.}(2021)\citenamefont
  {Calz\`a}, \citenamefont {March-Russell},\ and\ \citenamefont
  {Rosa}}]{Calza:2021czr}%
  \BibitemOpen
  \bibfield  {author} {\bibinfo {author} {\bibfnamefont {M.}~\bibnamefont
  {Calz\`a}}, \bibinfo {author} {\bibfnamefont {J.}~\bibnamefont
  {March-Russell}}, \ and\ \bibinfo {author} {\bibfnamefont {J.~a.~G.}\
  \bibnamefont {Rosa}},\ }\href@noop {} {\  (\bibinfo {year} {2021})},\ \Eprint
  {http://arxiv.org/abs/2110.13602} {arXiv:2110.13602 [astro-ph.CO]}
  \BibitemShut {NoStop}%
\bibitem [{\citenamefont {Schiavone}\ \emph {et~al.}(2021)\citenamefont
  {Schiavone}, \citenamefont {Montanino}, \citenamefont {Mirizzi},\ and\
  \citenamefont {Capozzi}}]{Schiavone:2021imu}%
  \BibitemOpen
  \bibfield  {author} {\bibinfo {author} {\bibfnamefont {F.}~\bibnamefont
  {Schiavone}}, \bibinfo {author} {\bibfnamefont {D.}~\bibnamefont
  {Montanino}}, \bibinfo {author} {\bibfnamefont {A.}~\bibnamefont {Mirizzi}},
  \ and\ \bibinfo {author} {\bibfnamefont {F.}~\bibnamefont {Capozzi}},\ }\href
  {\doibase 10.1088/1475-7516/2021/08/063} {\bibfield  {journal} {\bibinfo
  {journal} {JCAP}\ }\textbf {\bibinfo {volume} {08}},\ \bibinfo {pages} {063}
  (\bibinfo {year} {2021})},\ \Eprint {http://arxiv.org/abs/2107.03420}
  {arXiv:2107.03420 [hep-ph]} \BibitemShut {NoStop}%
\bibitem [{\citenamefont {Auffinger}(2022)}]{Auffinger:2022khh}%
  \BibitemOpen
  \bibfield  {author} {\bibinfo {author} {\bibfnamefont {J.}~\bibnamefont
  {Auffinger}},\ }\href@noop {} {\  (\bibinfo {year} {2022})},\ \Eprint
  {http://arxiv.org/abs/2206.02672} {arXiv:2206.02672 [astro-ph.CO]}
  \BibitemShut {NoStop}%
\bibitem [{\citenamefont {Kachulis}\ \emph {et~al.}(2018)\citenamefont
  {Kachulis} \emph {et~al.}}]{Super-Kamiokande:2017dch}%
  \BibitemOpen
  \bibfield  {author} {\bibinfo {author} {\bibfnamefont {C.}~\bibnamefont
  {Kachulis}} \emph {et~al.} (\bibinfo {collaboration} {Super-Kamiokande}),\
  }\href {\doibase 10.1103/PhysRevLett.120.221301} {\bibfield  {journal}
  {\bibinfo  {journal} {Phys. Rev. Lett.}\ }\textbf {\bibinfo {volume} {120}},\
  \bibinfo {pages} {221301} (\bibinfo {year} {2018})},\ \Eprint
  {http://arxiv.org/abs/1711.05278} {arXiv:1711.05278 [hep-ex]} \BibitemShut
  {NoStop}%
\bibitem [{\citenamefont {Ema}\ \emph {et~al.}(2019)\citenamefont {Ema},
  \citenamefont {Sala},\ and\ \citenamefont {Sato}}]{Ema:2018bih}%
  \BibitemOpen
  \bibfield  {author} {\bibinfo {author} {\bibfnamefont {Y.}~\bibnamefont
  {Ema}}, \bibinfo {author} {\bibfnamefont {F.}~\bibnamefont {Sala}}, \ and\
  \bibinfo {author} {\bibfnamefont {R.}~\bibnamefont {Sato}},\ }\href {\doibase
  10.1103/PhysRevLett.122.181802} {\bibfield  {journal} {\bibinfo  {journal}
  {Phys. Rev. Lett.}\ }\textbf {\bibinfo {volume} {122}},\ \bibinfo {pages}
  {181802} (\bibinfo {year} {2019})},\ \Eprint
  {http://arxiv.org/abs/1811.00520} {arXiv:1811.00520 [hep-ph]} \BibitemShut
  {NoStop}%
\bibitem [{\citenamefont {Lucente}\ \emph {et~al.}(2022)\citenamefont
  {Lucente}, \citenamefont {Mastrototaro}, \citenamefont {Carenza},
  \citenamefont {Di~Luzio}, \citenamefont {Giannotti},\ and\ \citenamefont
  {Mirizzi}}]{Lucente:2022vuo}%
  \BibitemOpen
  \bibfield  {author} {\bibinfo {author} {\bibfnamefont {G.}~\bibnamefont
  {Lucente}}, \bibinfo {author} {\bibfnamefont {L.}~\bibnamefont
  {Mastrototaro}}, \bibinfo {author} {\bibfnamefont {P.}~\bibnamefont
  {Carenza}}, \bibinfo {author} {\bibfnamefont {L.}~\bibnamefont {Di~Luzio}},
  \bibinfo {author} {\bibfnamefont {M.}~\bibnamefont {Giannotti}}, \ and\
  \bibinfo {author} {\bibfnamefont {A.}~\bibnamefont {Mirizzi}},\ }\href
  {\doibase 10.1103/PhysRevD.105.123020} {\bibfield  {journal} {\bibinfo
  {journal} {Phys. Rev. D}\ }\textbf {\bibinfo {volume} {105}},\ \bibinfo
  {pages} {123020} (\bibinfo {year} {2022})},\ \Eprint
  {http://arxiv.org/abs/2203.15812} {arXiv:2203.15812 [hep-ph]} \BibitemShut
  {NoStop}%
\bibitem [{\citenamefont {Abe}\ \emph {et~al.}(2018)\citenamefont {Abe} \emph
  {et~al.}}]{Hyper-Kamiokande:2018ofw}%
  \BibitemOpen
  \bibfield  {author} {\bibinfo {author} {\bibfnamefont {K.}~\bibnamefont
  {Abe}} \emph {et~al.} (\bibinfo {collaboration} {Hyper-Kamiokande}),\
  }\href@noop {} {\  (\bibinfo {year} {2018})},\ \Eprint
  {http://arxiv.org/abs/1805.04163} {arXiv:1805.04163 [physics.ins-det]}
  \BibitemShut {NoStop}%
\bibitem [{\citenamefont {Hawking}(1975)}]{Hawking:1975vcx}%
  \BibitemOpen
  \bibfield  {author} {\bibinfo {author} {\bibfnamefont {S.~W.}\ \bibnamefont
  {Hawking}},\ }\href {\doibase 10.1007/BF02345020} {\bibfield  {journal}
  {\bibinfo  {journal} {Commun. Math. Phys.}\ }\textbf {\bibinfo {volume}
  {43}},\ \bibinfo {pages} {199} (\bibinfo {year} {1975})},\ \bibinfo {note}
  {[Erratum: Commun.Math.Phys. 46, 206 (1976)]}\BibitemShut {NoStop}%
\bibitem [{\citenamefont {Page}(1976{\natexlab{a}})}]{Page:1976df}%
  \BibitemOpen
  \bibfield  {author} {\bibinfo {author} {\bibfnamefont {D.~N.}\ \bibnamefont
  {Page}},\ }\href {\doibase 10.1103/PhysRevD.13.198} {\bibfield  {journal}
  {\bibinfo  {journal} {Phys. Rev. D}\ }\textbf {\bibinfo {volume} {13}},\
  \bibinfo {pages} {198} (\bibinfo {year} {1976}{\natexlab{a}})}\BibitemShut
  {NoStop}%
\bibitem [{\citenamefont {Page}(1977)}]{Page:1977um}%
  \BibitemOpen
  \bibfield  {author} {\bibinfo {author} {\bibfnamefont {D.~N.}\ \bibnamefont
  {Page}},\ }\href {\doibase 10.1103/PhysRevD.16.2402} {\bibfield  {journal}
  {\bibinfo  {journal} {Phys. Rev. D}\ }\textbf {\bibinfo {volume} {16}},\
  \bibinfo {pages} {2402} (\bibinfo {year} {1977})}\BibitemShut {NoStop}%
\bibitem [{\citenamefont {MacGibbon}\ and\ \citenamefont
  {Webber}(1990)}]{MacGibbon:1990zk}%
  \BibitemOpen
  \bibfield  {author} {\bibinfo {author} {\bibfnamefont {J.~H.}\ \bibnamefont
  {MacGibbon}}\ and\ \bibinfo {author} {\bibfnamefont {B.~R.}\ \bibnamefont
  {Webber}},\ }\href {\doibase 10.1103/PhysRevD.41.3052} {\bibfield  {journal}
  {\bibinfo  {journal} {Phys. Rev. D}\ }\textbf {\bibinfo {volume} {41}},\
  \bibinfo {pages} {3052} (\bibinfo {year} {1990})}\BibitemShut {NoStop}%
\bibitem [{\citenamefont {MacGibbon}(1991)}]{MacGibbon:1991tj}%
  \BibitemOpen
  \bibfield  {author} {\bibinfo {author} {\bibfnamefont {J.~H.}\ \bibnamefont
  {MacGibbon}},\ }\href {\doibase 10.1103/PhysRevD.44.376} {\bibfield
  {journal} {\bibinfo  {journal} {Phys. Rev. D}\ }\textbf {\bibinfo {volume}
  {44}},\ \bibinfo {pages} {376} (\bibinfo {year} {1991})}\BibitemShut
  {NoStop}%
\bibitem [{\citenamefont {Hawking}(1971)}]{Hawking:1971ei}%
  \BibitemOpen
  \bibfield  {author} {\bibinfo {author} {\bibfnamefont {S.}~\bibnamefont
  {Hawking}},\ }\href@noop {} {\bibfield  {journal} {\bibinfo  {journal} {Mon.
  Not. Roy. Astron. Soc.}\ }\textbf {\bibinfo {volume} {152}},\ \bibinfo
  {pages} {75} (\bibinfo {year} {1971})}\BibitemShut {NoStop}%
\bibitem [{\citenamefont {Page}(1976{\natexlab{b}})}]{Page:1976ki}%
  \BibitemOpen
  \bibfield  {author} {\bibinfo {author} {\bibfnamefont {D.~N.}\ \bibnamefont
  {Page}},\ }\href {\doibase 10.1103/PhysRevD.14.3260} {\bibfield  {journal}
  {\bibinfo  {journal} {Phys. Rev. D}\ }\textbf {\bibinfo {volume} {14}},\
  \bibinfo {pages} {3260} (\bibinfo {year} {1976}{\natexlab{b}})}\BibitemShut
  {NoStop}%
\bibitem [{\citenamefont {Arbey}\ and\ \citenamefont
  {Auffinger}(2019)}]{Arbey:2019mbc}%
  \BibitemOpen
  \bibfield  {author} {\bibinfo {author} {\bibfnamefont {A.}~\bibnamefont
  {Arbey}}\ and\ \bibinfo {author} {\bibfnamefont {J.}~\bibnamefont
  {Auffinger}},\ }\href {\doibase 10.1140/epjc/s10052-019-7161-1} {\bibfield
  {journal} {\bibinfo  {journal} {Eur. Phys. J. C}\ }\textbf {\bibinfo {volume}
  {79}},\ \bibinfo {pages} {693} (\bibinfo {year} {2019})},\ \Eprint
  {http://arxiv.org/abs/1905.04268} {arXiv:1905.04268 [gr-qc]} \BibitemShut
  {NoStop}%
\bibitem [{\citenamefont {Arbey}\ and\ \citenamefont
  {Auffinger}(2021)}]{Arbey:2021mbl}%
  \BibitemOpen
  \bibfield  {author} {\bibinfo {author} {\bibfnamefont {A.}~\bibnamefont
  {Arbey}}\ and\ \bibinfo {author} {\bibfnamefont {J.}~\bibnamefont
  {Auffinger}},\ }\href {\doibase 10.1140/epjc/s10052-021-09702-8} {\bibfield
  {journal} {\bibinfo  {journal} {Eur. Phys. J. C}\ }\textbf {\bibinfo {volume}
  {81}},\ \bibinfo {pages} {910} (\bibinfo {year} {2021})},\ \Eprint
  {http://arxiv.org/abs/2108.02737} {arXiv:2108.02737 [gr-qc]} \BibitemShut
  {NoStop}%
\bibitem [{\citenamefont {Navarro}\ \emph {et~al.}(1997)\citenamefont
  {Navarro}, \citenamefont {Frenk},\ and\ \citenamefont
  {White}}]{Navarro:1996gj}%
  \BibitemOpen
  \bibfield  {author} {\bibinfo {author} {\bibfnamefont {J.~F.}\ \bibnamefont
  {Navarro}}, \bibinfo {author} {\bibfnamefont {C.~S.}\ \bibnamefont {Frenk}},
  \ and\ \bibinfo {author} {\bibfnamefont {S.~D.~M.}\ \bibnamefont {White}},\
  }\href {\doibase 10.1086/304888} {\bibfield  {journal} {\bibinfo  {journal}
  {Astrophys. J.}\ }\textbf {\bibinfo {volume} {490}},\ \bibinfo {pages} {493}
  (\bibinfo {year} {1997})},\ \Eprint {http://arxiv.org/abs/astro-ph/9611107}
  {arXiv:astro-ph/9611107} \BibitemShut {NoStop}%
\bibitem [{\citenamefont {Salucci}\ \emph {et~al.}(2010)\citenamefont
  {Salucci}, \citenamefont {Nesti}, \citenamefont {Gentile},\ and\
  \citenamefont {Martins}}]{Salucci:2010qr}%
  \BibitemOpen
  \bibfield  {author} {\bibinfo {author} {\bibfnamefont {P.}~\bibnamefont
  {Salucci}}, \bibinfo {author} {\bibfnamefont {F.}~\bibnamefont {Nesti}},
  \bibinfo {author} {\bibfnamefont {G.}~\bibnamefont {Gentile}}, \ and\
  \bibinfo {author} {\bibfnamefont {C.~F.}\ \bibnamefont {Martins}},\ }\href
  {\doibase 10.1051/0004-6361/201014385} {\bibfield  {journal} {\bibinfo
  {journal} {Astron. Astrophys.}\ }\textbf {\bibinfo {volume} {523}},\ \bibinfo
  {pages} {A83} (\bibinfo {year} {2010})},\ \Eprint
  {http://arxiv.org/abs/1003.3101} {arXiv:1003.3101 [astro-ph.GA]} \BibitemShut
  {NoStop}%
\bibitem [{\citenamefont {Yuan}\ \emph {et~al.}(2017)\citenamefont {Yuan},
  \citenamefont {Lin}, \citenamefont {Fang},\ and\ \citenamefont
  {Bi}}]{Yuan:2017ozr}%
  \BibitemOpen
  \bibfield  {author} {\bibinfo {author} {\bibfnamefont {Q.}~\bibnamefont
  {Yuan}}, \bibinfo {author} {\bibfnamefont {S.-J.}\ \bibnamefont {Lin}},
  \bibinfo {author} {\bibfnamefont {K.}~\bibnamefont {Fang}}, \ and\ \bibinfo
  {author} {\bibfnamefont {X.-J.}\ \bibnamefont {Bi}},\ }\href {\doibase
  10.1103/PhysRevD.95.083007} {\bibfield  {journal} {\bibinfo  {journal} {Phys.
  Rev. D}\ }\textbf {\bibinfo {volume} {95}},\ \bibinfo {pages} {083007}
  (\bibinfo {year} {2017})},\ \Eprint {http://arxiv.org/abs/1701.06149}
  {arXiv:1701.06149 [astro-ph.HE]} \BibitemShut {NoStop}%
\bibitem [{\citenamefont {Aghanim}\ \emph {et~al.}(2020)\citenamefont {Aghanim}
  \emph {et~al.}}]{Planck:2018vyg}%
  \BibitemOpen
  \bibfield  {author} {\bibinfo {author} {\bibfnamefont {N.}~\bibnamefont
  {Aghanim}} \emph {et~al.} (\bibinfo {collaboration} {Planck}),\ }\href
  {\doibase 10.1051/0004-6361/201833910} {\bibfield  {journal} {\bibinfo
  {journal} {Astron. Astrophys.}\ }\textbf {\bibinfo {volume} {641}},\ \bibinfo
  {pages} {A6} (\bibinfo {year} {2020})},\ \Eprint
  {http://arxiv.org/abs/1807.06209} {arXiv:1807.06209 [astro-ph.CO]}
  \BibitemShut {NoStop}%
\bibitem [{\citenamefont {Redondo}(2013)}]{Redondo:2013wwa}%
  \BibitemOpen
  \bibfield  {author} {\bibinfo {author} {\bibfnamefont {J.}~\bibnamefont
  {Redondo}},\ }\href {\doibase 10.1088/1475-7516/2013/12/008} {\bibfield
  {journal} {\bibinfo  {journal} {JCAP}\ }\textbf {\bibinfo {volume} {12}},\
  \bibinfo {pages} {008} (\bibinfo {year} {2013})},\ \Eprint
  {http://arxiv.org/abs/1310.0823} {arXiv:1310.0823 [hep-ph]} \BibitemShut
  {NoStop}%
\bibitem [{\citenamefont {Raffelt}(1986)}]{Raffelt:1985nk}%
  \BibitemOpen
  \bibfield  {author} {\bibinfo {author} {\bibfnamefont {G.~G.}\ \bibnamefont
  {Raffelt}},\ }\href {\doibase 10.1103/PhysRevD.33.897} {\bibfield  {journal}
  {\bibinfo  {journal} {Phys. Rev. D}\ }\textbf {\bibinfo {volume} {33}},\
  \bibinfo {pages} {897} (\bibinfo {year} {1986})}\BibitemShut {NoStop}%
\bibitem [{\citenamefont {Gondolo}\ and\ \citenamefont
  {Raffelt}(2009)}]{Gondolo:2008dd}%
  \BibitemOpen
  \bibfield  {author} {\bibinfo {author} {\bibfnamefont {P.}~\bibnamefont
  {Gondolo}}\ and\ \bibinfo {author} {\bibfnamefont {G.~G.}\ \bibnamefont
  {Raffelt}},\ }\href {\doibase 10.1103/PhysRevD.79.107301} {\bibfield
  {journal} {\bibinfo  {journal} {Phys. Rev. D}\ }\textbf {\bibinfo {volume}
  {79}},\ \bibinfo {pages} {107301} (\bibinfo {year} {2009})},\ \Eprint
  {http://arxiv.org/abs/0807.2926} {arXiv:0807.2926 [astro-ph]} \BibitemShut
  {NoStop}%
\bibitem [{\citenamefont {Caputo}\ \emph {et~al.}(2020)\citenamefont {Caputo},
  \citenamefont {Millar},\ and\ \citenamefont {Vitagliano}}]{Caputo:2020quz}%
  \BibitemOpen
  \bibfield  {author} {\bibinfo {author} {\bibfnamefont {A.}~\bibnamefont
  {Caputo}}, \bibinfo {author} {\bibfnamefont {A.~J.}\ \bibnamefont {Millar}},
  \ and\ \bibinfo {author} {\bibfnamefont {E.}~\bibnamefont {Vitagliano}},\
  }\href {\doibase 10.1103/PhysRevD.101.123004} {\bibfield  {journal} {\bibinfo
   {journal} {Phys. Rev. D}\ }\textbf {\bibinfo {volume} {101}},\ \bibinfo
  {pages} {123004} (\bibinfo {year} {2020})},\ \Eprint
  {http://arxiv.org/abs/2005.00078} {arXiv:2005.00078 [hep-ph]} \BibitemShut
  {NoStop}%
\bibitem [{\citenamefont {Carenza}\ and\ \citenamefont
  {Lucente}(2021)}]{Carenza:2021osu}%
  \BibitemOpen
  \bibfield  {author} {\bibinfo {author} {\bibfnamefont {P.}~\bibnamefont
  {Carenza}}\ and\ \bibinfo {author} {\bibfnamefont {G.}~\bibnamefont
  {Lucente}},\ }\href {\doibase 10.1103/PhysRevD.103.123024} {\bibfield
  {journal} {\bibinfo  {journal} {Phys. Rev. D}\ }\textbf {\bibinfo {volume}
  {103}},\ \bibinfo {pages} {123024} (\bibinfo {year} {2021})},\ \Eprint
  {http://arxiv.org/abs/2104.09524} {arXiv:2104.09524 [hep-ph]} \BibitemShut
  {NoStop}%
\bibitem [{\citenamefont {Giannotti}\ \emph {et~al.}(2017)\citenamefont
  {Giannotti}, \citenamefont {Irastorza}, \citenamefont {Redondo},
  \citenamefont {Ringwald},\ and\ \citenamefont {Saikawa}}]{Giannotti:2017hny}%
  \BibitemOpen
  \bibfield  {author} {\bibinfo {author} {\bibfnamefont {M.}~\bibnamefont
  {Giannotti}}, \bibinfo {author} {\bibfnamefont {I.~G.}\ \bibnamefont
  {Irastorza}}, \bibinfo {author} {\bibfnamefont {J.}~\bibnamefont {Redondo}},
  \bibinfo {author} {\bibfnamefont {A.}~\bibnamefont {Ringwald}}, \ and\
  \bibinfo {author} {\bibfnamefont {K.}~\bibnamefont {Saikawa}},\ }\href
  {\doibase 10.1088/1475-7516/2017/10/010} {\bibfield  {journal} {\bibinfo
  {journal} {JCAP}\ }\textbf {\bibinfo {volume} {10}},\ \bibinfo {pages} {010}
  (\bibinfo {year} {2017})},\ \Eprint {http://arxiv.org/abs/1708.02111}
  {arXiv:1708.02111 [hep-ph]} \BibitemShut {NoStop}%
\bibitem [{\citenamefont {Aprile}\ \emph {et~al.}(2020)\citenamefont {Aprile}
  \emph {et~al.}}]{XENON:2020rca}%
  \BibitemOpen
  \bibfield  {author} {\bibinfo {author} {\bibfnamefont {E.}~\bibnamefont
  {Aprile}} \emph {et~al.} (\bibinfo {collaboration} {XENON}),\ }\href
  {\doibase 10.1103/PhysRevD.102.072004} {\bibfield  {journal} {\bibinfo
  {journal} {Phys. Rev. D}\ }\textbf {\bibinfo {volume} {102}},\ \bibinfo
  {pages} {072004} (\bibinfo {year} {2020})},\ \Eprint
  {http://arxiv.org/abs/2006.09721} {arXiv:2006.09721 [hep-ex]} \BibitemShut
  {NoStop}%
\bibitem [{\citenamefont {Pospelov}\ \emph {et~al.}(2008)\citenamefont
  {Pospelov}, \citenamefont {Ritz},\ and\ \citenamefont
  {Voloshin}}]{Pospelov:2008jk}%
  \BibitemOpen
  \bibfield  {author} {\bibinfo {author} {\bibfnamefont {M.}~\bibnamefont
  {Pospelov}}, \bibinfo {author} {\bibfnamefont {A.}~\bibnamefont {Ritz}}, \
  and\ \bibinfo {author} {\bibfnamefont {M.~B.}\ \bibnamefont {Voloshin}},\
  }\href {\doibase 10.1103/PhysRevD.78.115012} {\bibfield  {journal} {\bibinfo
  {journal} {Phys. Rev. D}\ }\textbf {\bibinfo {volume} {78}},\ \bibinfo
  {pages} {115012} (\bibinfo {year} {2008})},\ \Eprint
  {http://arxiv.org/abs/0807.3279} {arXiv:0807.3279 [hep-ph]} \BibitemShut
  {NoStop}%
\bibitem [{\citenamefont {Bloch}\ \emph {et~al.}(2017)\citenamefont {Bloch},
  \citenamefont {Essig}, \citenamefont {Tobioka}, \citenamefont {Volansky},\
  and\ \citenamefont {Yu}}]{Bloch:2016sjj}%
  \BibitemOpen
  \bibfield  {author} {\bibinfo {author} {\bibfnamefont {I.~M.}\ \bibnamefont
  {Bloch}}, \bibinfo {author} {\bibfnamefont {R.}~\bibnamefont {Essig}},
  \bibinfo {author} {\bibfnamefont {K.}~\bibnamefont {Tobioka}}, \bibinfo
  {author} {\bibfnamefont {T.}~\bibnamefont {Volansky}}, \ and\ \bibinfo
  {author} {\bibfnamefont {T.-T.}\ \bibnamefont {Yu}},\ }\href {\doibase
  10.1007/JHEP06(2017)087} {\bibfield  {journal} {\bibinfo  {journal} {JHEP}\
  }\textbf {\bibinfo {volume} {06}},\ \bibinfo {pages} {087} (\bibinfo {year}
  {2017})},\ \Eprint {http://arxiv.org/abs/1608.02123} {arXiv:1608.02123
  [hep-ph]} \BibitemShut {NoStop}%
\bibitem [{\citenamefont {Henke}\ \emph {et~al.}(1993)\citenamefont {Henke},
  \citenamefont {Gullikson},\ and\ \citenamefont {Davis}}]{Henke:1993eda}%
  \BibitemOpen
  \bibfield  {author} {\bibinfo {author} {\bibfnamefont {B.~L.}\ \bibnamefont
  {Henke}}, \bibinfo {author} {\bibfnamefont {E.~M.}\ \bibnamefont
  {Gullikson}}, \ and\ \bibinfo {author} {\bibfnamefont {J.~C.}\ \bibnamefont
  {Davis}},\ }\href {\doibase 10.1006/adnd.1993.1013} {\bibfield  {journal}
  {\bibinfo  {journal} {Atom. Data Nucl. Data Tabl.}\ }\textbf {\bibinfo
  {volume} {54}},\ \bibinfo {pages} {181} (\bibinfo {year} {1993})}\BibitemShut
  {NoStop}%
\bibitem [{\citenamefont {Gao}\ \emph {et~al.}(2020)\citenamefont {Gao},
  \citenamefont {Liu}, \citenamefont {Wang}, \citenamefont {Wang},
  \citenamefont {Xue},\ and\ \citenamefont {Zhong}}]{Gao:2020wer}%
  \BibitemOpen
  \bibfield  {author} {\bibinfo {author} {\bibfnamefont {C.}~\bibnamefont
  {Gao}}, \bibinfo {author} {\bibfnamefont {J.}~\bibnamefont {Liu}}, \bibinfo
  {author} {\bibfnamefont {L.-T.}\ \bibnamefont {Wang}}, \bibinfo {author}
  {\bibfnamefont {X.-P.}\ \bibnamefont {Wang}}, \bibinfo {author}
  {\bibfnamefont {W.}~\bibnamefont {Xue}}, \ and\ \bibinfo {author}
  {\bibfnamefont {Y.-M.}\ \bibnamefont {Zhong}},\ }\href {\doibase
  10.1103/PhysRevLett.125.131806} {\bibfield  {journal} {\bibinfo  {journal}
  {Phys. Rev. Lett.}\ }\textbf {\bibinfo {volume} {125}},\ \bibinfo {pages}
  {131806} (\bibinfo {year} {2020})},\ \Eprint
  {http://arxiv.org/abs/2006.14598} {arXiv:2006.14598 [hep-ph]} \BibitemShut
  {NoStop}%
\bibitem [{\citenamefont {Dent}\ \emph {et~al.}(2020)\citenamefont {Dent},
  \citenamefont {Dutta}, \citenamefont {Newstead},\ and\ \citenamefont
  {Thompson}}]{Dent:2020jhf}%
  \BibitemOpen
  \bibfield  {author} {\bibinfo {author} {\bibfnamefont {J.~B.}\ \bibnamefont
  {Dent}}, \bibinfo {author} {\bibfnamefont {B.}~\bibnamefont {Dutta}},
  \bibinfo {author} {\bibfnamefont {J.~L.}\ \bibnamefont {Newstead}}, \ and\
  \bibinfo {author} {\bibfnamefont {A.}~\bibnamefont {Thompson}},\ }\href
  {\doibase 10.1103/PhysRevLett.125.131805} {\bibfield  {journal} {\bibinfo
  {journal} {Phys. Rev. Lett.}\ }\textbf {\bibinfo {volume} {125}},\ \bibinfo
  {pages} {131805} (\bibinfo {year} {2020})},\ \Eprint
  {http://arxiv.org/abs/2006.15118} {arXiv:2006.15118 [hep-ph]} \BibitemShut
  {NoStop}%
\bibitem [{\citenamefont {Bloch}\ \emph {et~al.}(2021)\citenamefont {Bloch},
  \citenamefont {Caputo}, \citenamefont {Essig}, \citenamefont {Redigolo},
  \citenamefont {Sholapurkar},\ and\ \citenamefont {Volansky}}]{Bloch:2020uzh}%
  \BibitemOpen
  \bibfield  {author} {\bibinfo {author} {\bibfnamefont {I.~M.}\ \bibnamefont
  {Bloch}}, \bibinfo {author} {\bibfnamefont {A.}~\bibnamefont {Caputo}},
  \bibinfo {author} {\bibfnamefont {R.}~\bibnamefont {Essig}}, \bibinfo
  {author} {\bibfnamefont {D.}~\bibnamefont {Redigolo}}, \bibinfo {author}
  {\bibfnamefont {M.}~\bibnamefont {Sholapurkar}}, \ and\ \bibinfo {author}
  {\bibfnamefont {T.}~\bibnamefont {Volansky}},\ }\href {\doibase
  10.1007/JHEP01(2021)178} {\bibfield  {journal} {\bibinfo  {journal} {JHEP}\
  }\textbf {\bibinfo {volume} {01}},\ \bibinfo {pages} {178} (\bibinfo {year}
  {2021})},\ \Eprint {http://arxiv.org/abs/2006.14521} {arXiv:2006.14521
  [hep-ph]} \BibitemShut {NoStop}%
\bibitem [{\citenamefont {Di~Luzio}\ \emph
  {et~al.}(2020{\natexlab{b}})\citenamefont {Di~Luzio}, \citenamefont {Fedele},
  \citenamefont {Giannotti}, \citenamefont {Mescia},\ and\ \citenamefont
  {Nardi}}]{DiLuzio:2020jjp}%
  \BibitemOpen
  \bibfield  {author} {\bibinfo {author} {\bibfnamefont {L.}~\bibnamefont
  {Di~Luzio}}, \bibinfo {author} {\bibfnamefont {M.}~\bibnamefont {Fedele}},
  \bibinfo {author} {\bibfnamefont {M.}~\bibnamefont {Giannotti}}, \bibinfo
  {author} {\bibfnamefont {F.}~\bibnamefont {Mescia}}, \ and\ \bibinfo {author}
  {\bibfnamefont {E.}~\bibnamefont {Nardi}},\ }\href {\doibase
  10.1103/PhysRevLett.125.131804} {\bibfield  {journal} {\bibinfo  {journal}
  {Phys. Rev. Lett.}\ }\textbf {\bibinfo {volume} {125}},\ \bibinfo {pages}
  {131804} (\bibinfo {year} {2020}{\natexlab{b}})},\ \Eprint
  {http://arxiv.org/abs/2006.12487} {arXiv:2006.12487 [hep-ph]} \BibitemShut
  {NoStop}%
\bibitem [{\citenamefont {Abe}\ \emph {et~al.}(2021)\citenamefont {Abe},
  \citenamefont {Hamaguchi},\ and\ \citenamefont {Nagata}}]{Abe:2020mcs}%
  \BibitemOpen
  \bibfield  {author} {\bibinfo {author} {\bibfnamefont {T.}~\bibnamefont
  {Abe}}, \bibinfo {author} {\bibfnamefont {K.}~\bibnamefont {Hamaguchi}}, \
  and\ \bibinfo {author} {\bibfnamefont {N.}~\bibnamefont {Nagata}},\ }\href
  {\doibase 10.1016/j.physletb.2021.136174} {\bibfield  {journal} {\bibinfo
  {journal} {Phys. Lett. B}\ }\textbf {\bibinfo {volume} {815}},\ \bibinfo
  {pages} {136174} (\bibinfo {year} {2021})},\ \Eprint
  {http://arxiv.org/abs/2012.02508} {arXiv:2012.02508 [hep-ph]} \BibitemShut
  {NoStop}%
\bibitem [{\citenamefont {Creswick}\ \emph {et~al.}(1998)\citenamefont
  {Creswick}, \citenamefont {Avignone}, \citenamefont {Farach}, \citenamefont
  {Collar}, \citenamefont {Gattone}, \citenamefont {Nussinov},\ and\
  \citenamefont {Zioutas}}]{Creswick:1997pg}%
  \BibitemOpen
  \bibfield  {author} {\bibinfo {author} {\bibfnamefont {R.~J.}\ \bibnamefont
  {Creswick}}, \bibinfo {author} {\bibfnamefont {F.~T.}\ \bibnamefont
  {Avignone}, \bibfnamefont {III}}, \bibinfo {author} {\bibfnamefont {H.~A.}\
  \bibnamefont {Farach}}, \bibinfo {author} {\bibfnamefont {J.~I.}\
  \bibnamefont {Collar}}, \bibinfo {author} {\bibfnamefont {A.~O.}\
  \bibnamefont {Gattone}}, \bibinfo {author} {\bibfnamefont {S.}~\bibnamefont
  {Nussinov}}, \ and\ \bibinfo {author} {\bibfnamefont {K.}~\bibnamefont
  {Zioutas}},\ }\href {\doibase 10.1016/S0370-2693(98)00183-X} {\bibfield
  {journal} {\bibinfo  {journal} {Phys. Lett. B}\ }\textbf {\bibinfo {volume}
  {427}},\ \bibinfo {pages} {235} (\bibinfo {year} {1998})},\ \Eprint
  {http://arxiv.org/abs/hep-ph/9708210} {arXiv:hep-ph/9708210} \BibitemShut
  {NoStop}%
\bibitem [{\citenamefont {Emken}\ and\ \citenamefont
  {Kouvaris}(2018)}]{Emken:2018run}%
  \BibitemOpen
  \bibfield  {author} {\bibinfo {author} {\bibfnamefont {T.}~\bibnamefont
  {Emken}}\ and\ \bibinfo {author} {\bibfnamefont {C.}~\bibnamefont
  {Kouvaris}},\ }\href {\doibase 10.1103/PhysRevD.97.115047} {\bibfield
  {journal} {\bibinfo  {journal} {Phys. Rev. D}\ }\textbf {\bibinfo {volume}
  {97}},\ \bibinfo {pages} {115047} (\bibinfo {year} {2018})},\ \Eprint
  {http://arxiv.org/abs/1802.04764} {arXiv:1802.04764 [hep-ph]} \BibitemShut
  {NoStop}%
\bibitem [{\citenamefont {Caputo}\ \emph
  {et~al.}(2022{\natexlab{a}})\citenamefont {Caputo}, \citenamefont {Janka},
  \citenamefont {Raffelt},\ and\ \citenamefont {Vitagliano}}]{Caputo:2022mah}%
  \BibitemOpen
  \bibfield  {author} {\bibinfo {author} {\bibfnamefont {A.}~\bibnamefont
  {Caputo}}, \bibinfo {author} {\bibfnamefont {H.-T.}\ \bibnamefont {Janka}},
  \bibinfo {author} {\bibfnamefont {G.}~\bibnamefont {Raffelt}}, \ and\
  \bibinfo {author} {\bibfnamefont {E.}~\bibnamefont {Vitagliano}},\ }\href
  {\doibase 10.1103/PhysRevLett.128.221103} {\bibfield  {journal} {\bibinfo
  {journal} {Phys. Rev. Lett.}\ }\textbf {\bibinfo {volume} {128}},\ \bibinfo
  {pages} {221103} (\bibinfo {year} {2022}{\natexlab{a}})},\ \Eprint
  {http://arxiv.org/abs/2201.09890} {arXiv:2201.09890 [astro-ph.HE]}
  \BibitemShut {NoStop}%
\bibitem [{\citenamefont {Dolan}\ \emph {et~al.}(2017)\citenamefont {Dolan},
  \citenamefont {Ferber}, \citenamefont {Hearty}, \citenamefont {Kahlhoefer},\
  and\ \citenamefont {Schmidt-Hoberg}}]{Dolan_2017}%
  \BibitemOpen
  \bibfield  {author} {\bibinfo {author} {\bibfnamefont {M.~J.}\ \bibnamefont
  {Dolan}}, \bibinfo {author} {\bibfnamefont {T.}~\bibnamefont {Ferber}},
  \bibinfo {author} {\bibfnamefont {C.}~\bibnamefont {Hearty}}, \bibinfo
  {author} {\bibfnamefont {F.}~\bibnamefont {Kahlhoefer}}, \ and\ \bibinfo
  {author} {\bibfnamefont {K.}~\bibnamefont {Schmidt-Hoberg}},\ }\href
  {\doibase 10.1007/jhep12(2017)094} {\bibfield  {journal} {\bibinfo  {journal}
  {Journal of High Energy Physics}\ }\textbf {\bibinfo {volume} {2017}}
  (\bibinfo {year} {2017}),\ 10.1007/jhep12(2017)094}\BibitemShut {NoStop}%
\bibitem [{\citenamefont {Caputo}\ \emph
  {et~al.}(2022{\natexlab{b}})\citenamefont {Caputo}, \citenamefont {Raffelt},\
  and\ \citenamefont {Vitagliano}}]{Caputo_2022}%
  \BibitemOpen
  \bibfield  {author} {\bibinfo {author} {\bibfnamefont {A.}~\bibnamefont
  {Caputo}}, \bibinfo {author} {\bibfnamefont {G.}~\bibnamefont {Raffelt}}, \
  and\ \bibinfo {author} {\bibfnamefont {E.}~\bibnamefont {Vitagliano}},\
  }\href {\doibase 10.1103/physrevd.105.035022} {\bibfield  {journal} {\bibinfo
   {journal} {Physical Review D}\ }\textbf {\bibinfo {volume} {105}} (\bibinfo
  {year} {2022}{\natexlab{b}}),\ 10.1103/physrevd.105.035022}\BibitemShut
  {NoStop}%
\bibitem [{\citenamefont {Jaeckel}\ \emph {et~al.}(2018)\citenamefont
  {Jaeckel}, \citenamefont {Malta},\ and\ \citenamefont
  {Redondo}}]{Jaeckel_2018}%
  \BibitemOpen
  \bibfield  {author} {\bibinfo {author} {\bibfnamefont {J.}~\bibnamefont
  {Jaeckel}}, \bibinfo {author} {\bibfnamefont {P.}~\bibnamefont {Malta}}, \
  and\ \bibinfo {author} {\bibfnamefont {J.}~\bibnamefont {Redondo}},\ }\href
  {\doibase 10.1103/physrevd.98.055032} {\bibfield  {journal} {\bibinfo
  {journal} {Physical Review D}\ }\textbf {\bibinfo {volume} {98}} (\bibinfo
  {year} {2018}),\ 10.1103/physrevd.98.055032}\BibitemShut {NoStop}%
\bibitem [{\citenamefont {Cui}\ \emph {et~al.}(2022)\citenamefont {Cui},
  \citenamefont {Kuo}, \citenamefont {Pradler},\ and\ \citenamefont
  {Tsai}}]{Cui:2022owf}%
  \BibitemOpen
  \bibfield  {author} {\bibinfo {author} {\bibfnamefont {Y.}~\bibnamefont
  {Cui}}, \bibinfo {author} {\bibfnamefont {J.-L.}\ \bibnamefont {Kuo}},
  \bibinfo {author} {\bibfnamefont {J.}~\bibnamefont {Pradler}}, \ and\
  \bibinfo {author} {\bibfnamefont {Y.-D.}\ \bibnamefont {Tsai}},\ }\href@noop
  {} {\  (\bibinfo {year} {2022})},\ \Eprint {http://arxiv.org/abs/2207.13107}
  {arXiv:2207.13107 [hep-ph]} \BibitemShut {NoStop}%
\end{thebibliography}%

\end{document}